\documentclass[11pt]{article}%
\usepackage{amsmath}
\usepackage{amsthm}
\usepackage{dsfont}
\usepackage{graphicx}
\usepackage{mathdots}
\usepackage{xcolor}%
\usepackage{amsfonts}%
\usepackage{amssymb}
\newcommand{\eqnb}{\begin{equation}}
\newcommand{\eqne}{\end{equation}}

\newtheorem{The}{Theorem}

\newtheorem{Cor}[The]{Corollary}
\newtheorem{Lem}{Lemma}

\newtheorem{Rem}{Remark}

\begin{document}

\title{\textbf{Two Basic Queueing Models of Service Platforms in Digital Sharing Economy}}
\author{Heng-Li Liu$^{a}$, Quan-Lin Li$^{b,}$\thanks{Corresponding author: Q.L. Li
(liquanlin@tsinghua.edu.cn)}, Xiaole Wu$^{c}$, Chi Zhang$^{b}$\\$^{a}$School of Economics and Management Sciences \\Yanshan University, Qinhuangdao 066004, China\\$^{b}$School of Economics and Management \\Beijing University of Technology, Beijing 100124, China \\$^{c}$School of Management \\Fudan University, Shanghai 200433, China}
\maketitle

\begin{abstract}
This paper describes two basic queueing models of service platforms in digital
sharing economy by means of two different policies of platform matching
information. We show that the two queueing models of service platforms can be
expressed as the level-independent quasi birth-and-death (QBD) processes.
Using the proposed QBD processes, we provide a detailed analysis for the two
queueing models of service platforms, including the system stability, the
average stationary numbers of seekers and of idle owners, the expected sojourn
time of an arriving seeker, and the expected profits for both the service
platform and each owner. Finally, numerical examples are employed to verify
our theoretical results, and demonstrate how the performance measures of
service platforms are influenced by some key system parameters. We believe
that the methodology and results developed in this paper not only can be
applied to develop a broad class of queuing models of service platforms, but
also will open a series of promising innovative research on performance
evaluation, optimal control and queueing-game of service platforms and digital
sharing economy.


\end{abstract}

\section{Introduction}

In the past few years, with great advances of wireless, mobile, Internet and
digital technologies, we have witnessed rapid rise of digital, platform and
sharing economies, which have become a class of increasingly important
economic modes in the current world. Many famous companies of service
platforms continue to emerge and develop rapidly. Important examples include
\textit{taxi-style transportation}: Uber, Lyft, Didi; \textit{rental housing}:
AirBnB, HomeAway; \textit{restaurant food delivery}: Caviar, DoorDash;
\textit{consumer goods delivery}: UberRush, Go-Mart; and so forth. For more
details of sharing economy, readers may refer to survey papers by Narasimhan
et al. \cite{Nar:2018}, Agarwal and Steinmetz \cite{Aga:2019}, Hossain
\cite{Hos:2020} and Kraus et al. \cite{Kra:2020}.

A service platform connects \textit{service requirements}, called seekers
(e.g., subscribers, customers) with \textit{service providers}, called owners
(e.g., contractors, suppliers, agents, taxi drivers). An owner receives a
payment from the service platform once a service is completed. The owners are
mutually independent in the sense that each of them can separately decide
whether and when to work. The service platform can operate well with the
current digital and information technologies, and it provides a matching
structure in a bilateral or multilateral market. It is worthwhile to note that
the service platforms play a key role in the digital sharing economy. The
readers may refer to, for example, \textit{survey papers} by Breidbach and
Brodie \cite{Bre:2017}, Sutherland and Jarrahi \cite{Sut:2018} and Costello
and Reczek \cite{Cos:2020}; \textit{key research} by Wirtz et al.
\cite{Wir:2019}, Choi and He \cite{ChoiH:2019}, Clauss et al. \cite{Cla:2019},
Wen and Siqin \cite{Wen:2020}, Cachon et al. \cite{Cac:2017} and Kung and
Zhong \cite{Kun:2017}; and \textit{practical service platforms} include food
by Choi et al. \cite{Choi:2019}, ride-hailing by Feng et al. \cite{FengT:2020}%
, hotel by Akbar and Tracogna \cite{Akb:2018}, E-tailing by Cho et al.
\cite{Cho:2019} and Gong et al. \cite{Gon:2020}, and Airbnb by Leoni and
Parker \cite{Len:2019} and Xu et al. \cite{Xu:2021}.

It is an interesting topic to develop queuing models that can sufficiently
express basic characteristics and physical structure of the service platforms.
To this end, this paper makes necessary exploration how to design such queuing
models of service platforms (e.g., see Figure 1). It is worthwhile to note
that the matching processes play a key role in the study of service platforms.
To this end, for a matched (or double-ended) queue, readers may refer to, for
example, Adan et al. \cite{Ada:2018}, Weiss \cite{Wei:2020}, Castro et al.
\cite{Cas:2020}, Liu et al. \cite{LiuH:2020, LiuHL:2020} and the references
therein. It is a key in queueing analysis of {service platforms} that we find
two different policies of platform matching information. \textit{Policy one}:
The platform matching information is over at the moment that the matching of a
seeker and an owner is completed and their service begins thereafter
immediately. Policy one motivates us to set up the first queuing model of
service platforms, which has not been studied in the literature up to now yet.
Note that Policy one can be well related to familiar {service platforms in
matching the }ordinary items from the bilateral markets, for example,
taxi-style transportation, and medical appointment. \textit{Policy two}: The
platform matching information is over at the time that a seeker and an owner
start a matching process. In this case, the matching and service processes are
completed sequentially. This leads to our second queuing model, corresponding
to the PH/PH/$N$ queue. Note that Policy two can be used to high-value
{service platforms in matching the }precious items from the bilateral markets,
for instance, jewelry reservation, and forward sale of houses. In this paper,
our two queuing systems of service platforms try to keep simple. Therefore,
both of them can be generalized from different perspectives, such that we can
open a series of promising innovative research on performance evaluation,
optimal control and queueing-game of service platforms and digital sharing economy.

So far little work has been done on the queueing analysis of service platforms
in digital sharing economy. Now, we review few recent literature on the
queueing models of service platforms from several perspectives. Kim and Yeun
\cite{Kim:2019} proposed a G$^{X}/$M$/1$ type queue to describe and analyze
the sharing economy platforms. Wang and Yan \cite{Wang:2019} used the
M/M/$1$/$K$ queue to describe a taxi--passenger dispatching model, the pairs
of which are matched between the two queues of taxis and passengers. Li and
Fan \cite{Li:2021} applied the mean-field theory to set up a MAP$_{t}$%
/PH$_{t}$/$1$ queue in the bike-sharing system with a Markovian environment.
From the above analysis, it is easy to see that those works only applied the
known queueing systems to express the service platforms (or sharing systems),
but they did not find a class of new queuing systems which have the
characteristics and context of service platforms. This motivates us in this
paper to find two new queueing models of service platforms.

Some studies have applied the queueing theory to the pricing control of
service platforms. Readers may refer to recent publications for details, among
which Cachon and Feldman \cite{Cac:2011} applied the queueing theory to find
that a firm may prefer to subscription pricing over per-use pricing even if
consumers dislike congestion. Banerjee et al. \cite{Ban:2015} set up a
queueing economic model to study the optimal pricing of a ride-sharing
platform. Banerjee et al. \cite{Ban:2016} presented a formal framework for
point-to-point pricing in a closed queueing network, and then analyzed the
vehicle sharing systems. Taylor \cite{Tay:2018} used an M/M/$n$ queue to
discuss an on-demand service platform and analyzed the service platform's
price and wage decisions. Bai et al. \cite{Bai2019} proposed a queueing model
that considers the earnings-sensitive independent drivers with heterogeneous
reservation prices, and the price-sensitive passengers with heterogeneous
valuations of the service. Zhong et al. \cite{Zhong:2019} compared surge
pricing and static pricing queues from multiple perspectives. Although the
above mentioned studies touched the pricing control issues of service
platforms, the queueing models they adopted are not consistent with the two
policies of platform matching information we find, and, thus, are not
practical. It is interesting to develop more practical queueing models to
study the pricing control issue of service platforms through using our two
queueing models and their generalization.

It is important to note that the matching resources of a service platform are
always scattered among different geographical locations. Thus, it is necessary
and interesting to discuss spatial queues of service platforms in digital
sharing economy. For earlier research on spatial queues with finer
granularity, readers may refer to Bertsimas and Ryfin \cite{Bert:1991},
Bertsimas and Ryfin \cite{Bert:1993} and Serfozo \cite{Ser:1999}. Chu et al.
\cite{Chu:2018} studied a single-location model in which drivers can
cherry-pick riders, and focused on information, routing, and priority controls
by the service platform. Af\`{e}che et al. \cite{Afe:2018} considered two
locations and focused on the performance impact of drivers' self-repositioning
and demand-side admission control. Braverman et al. \cite{Bra:2019} discussed
multiple locations and focused on empty-car routing control. Besbes et al.
\cite{Bes:2018} provided an M/M/$k$ queue with a state-dependent service rate
that takes into account the pickup time under the match-to-the-closest
dispatch rule. Feng et al. \cite{Feng:2020} examined the on-demand hailing and
traditional street-hailing systems by using an M/M/$k$ queueing approximation.
Hu \cite{Hu:2020} adopted the queuing theory to capture the spatial movements
of vehicles in a centralized vehicle-sharing system. Chen and Hu
\cite{Chen:2020} analyzed a courier dispatching problem in an on-demand
delivery system where customers are sensitive to delay.\ He et al.
\cite{He:2019} proposed an integrated model of service platforms to understand
the operations of shared-mobility systems. Sun et al. \cite{Sun:2020} used an
approximate queue to explore how the destination preference affects a driver's
system choice, and how the service platform optimally allocated rides to both
system structure and setting of radius. From the above discussions, similarly
with those studies on pricing control reviewed in the previous paragraph, the
research on spatial queues also employed queueing models that are inconsistent
with the two policies of platform matching information we find, such that they
are not also practical. Therefore, it is an interesting topic to develop new
practically spatial queueing models to be able to effectively support the
matching and service processes of service platforms.

The main contributions of this paper can be summarized in three-fold.

\begin{itemize}
\item[(1)] We describe two basic queueing models of service platforms
according to two different policies of platform matching information: One is
at the matching completion time (it is also the service beginning time), while
another is at the matching beginning time (no service yet). In addition, we
inductively find several practical factors: (a) finite number of owners, (b)
infinite number of seekers, (c) the matching process derived from the fact
that an owner can match a seeker as a pair, (d) the service process for each
pair of matched owner and seeker, and (e) the phenomenon that once the service
is completed, the owner returns to the service platform, while the seeker
immediately leaves the system. See Figure 1 for more details. This queueing
model captures the basic characteristics and physical structure of service
platforms, and can motivate a series of new interesting queueing systems of
service platforms.

\item[(2)] We express the two basic queueing models of service platforms as
the level-independent QBD processes, and apply the matrix-geometric method to
obtain a necessary and sufficient condition under which the system is stable,
the stationary probability vector, the expected sojourn time by using the
RG-factorizations, and the expected profits for the service platform and each
owner. This enables the performance analysis of service platforms.

\item[(3)] We use some numerical examples to illustrate our theoretical
results, and show how performance measures of service platforms are influenced
by the identified key system parameters.
\end{itemize}

The structure of this paper is organized as follows. Section 2 describes two
basic queueing models of service platforms in digital sharing economy.
Sections 3 and 4 analyze the first queueing model of service platforms.
Section 3 expresses the first model as a level-independent QBD process, and
obtains a necessary and sufficient condition under which the system is stable.
Section 4 applies the matrix-geometric solution to derive the stationary
probability vector of the QBD process, and then provides some useful
performance measures of the service platforms. Section 5 simply analyzes the
second queueing model of service platforms by means of another QBD process.
Section 6 uses numerical examples to discuss how the performance measures
depend on some key system parameters. Section 7 provides concluding remarks.

\section{Model Description}

In this section, we describe two basic queueing models of service platforms in
digital sharing economy, and provide some necessary notation used in our
following analysis.

For a service platform in digital sharing economy, we need to capture the
following factors: owners, seekers, the matching process between owners and
seekers, the service process for the matched pairs of owners and seekers, and
the fact that the owner returns to the service platform and the seeker
immediately leaves the system once the service is completed. In addition, we
need to determine how a service price is paid by each seeker, and how the
expected profits are allocated between the service platform and each owner.

It is a key to find two policies of platform matching information. This
motivates us to design two basic queueing models of service platforms as follows:

\textbf{Model one:} The information completed in a service platform is
observed at the matching completion time (it is also the service beginning time).

\textbf{Model two:} The information completed in a service platform is
observed at the matching beginning time (no service yet). In this case, the
matching and service times may be regarded as forming a generalized service
time, i.e., the sum of a matching time and a service time.

We adopt the following assumptions for the two queueing models:

\textbf{(1) The owners:} There are $N$ independent owners who have registered
in the service platform, i.e., all the service resources of the platform are
the $N$ owners. When an owner is serving a seeker, he/she cannot accept any
new task assignments. Once the owner completes his/her service for the seeker,
he/she immediately returns to the resource of the service platform and waits
for a new task assignment.

\textbf{(2) The seekers:} The service requirements of the seekers arrive at
the service platform according to a Poisson process with rate $\lambda$. A
seeker can successfully submit only one service requirement to the service
platform. Meanwhile, the service platform can receive an infinite number of
service requirements from the seekers.

\textbf{(3) The matching process:} If there is at least one idle owner who has
no service task arranged by the service platform, and the service requirement
of a seeker arrives at the service platform, then the idle owner and the
seeker begin to match as a pair, and the matching time of the seeker and an
owner is exponential with matching rate $\gamma$, i.e., the average matching
time is $1/\gamma$.

To match the seeker, one owner is chosen based on a number of practical
factors, such as the reputation ratings of owners, the matching distance, the
matching difficulty, and so forth. To some extent, the matching rate $\gamma$
reflects a comprehensive effect of these practical factors.

\textbf{(4) The service process:} Once a seeker is matched with an owner as a
pair, he/she is immediately served from the owner. The service time of each
matching pair of a seeker and an owner is exponential with service rate $\mu$,
i.e., the average service time is $1/\mu$. Note that a working owner cannot
receive any new task assignment for the seekers.

Once the service of a matching pair is completed, the owner returns to and
becomes available in the service platform, while the seeker leaves the system
immediately. Note that the seekers are served based on the First Come First
Served (FCFS) discipline.

\textbf{(5) The service price:} $P$ is the service price charged to each
seeker given that the required service is completed.

\textbf{(6) The profit allocation proportion:} For the service price $P$, once
the owner completes the service of the seeker, the proportion $d$ of the
service price $P$ is paid to the owner, while the rest of the price is
retained by the service platform, where, $0<d<1$.

\textbf{(7) The independence:} We assume that all the random variables defined
above, such as the matching and service times, are independent of each other.

Figure 1 is an illustration of the queueing structure of service platforms.

\begin{figure}[tbh]
\centering       \includegraphics[height=7cm
,width=12cm]{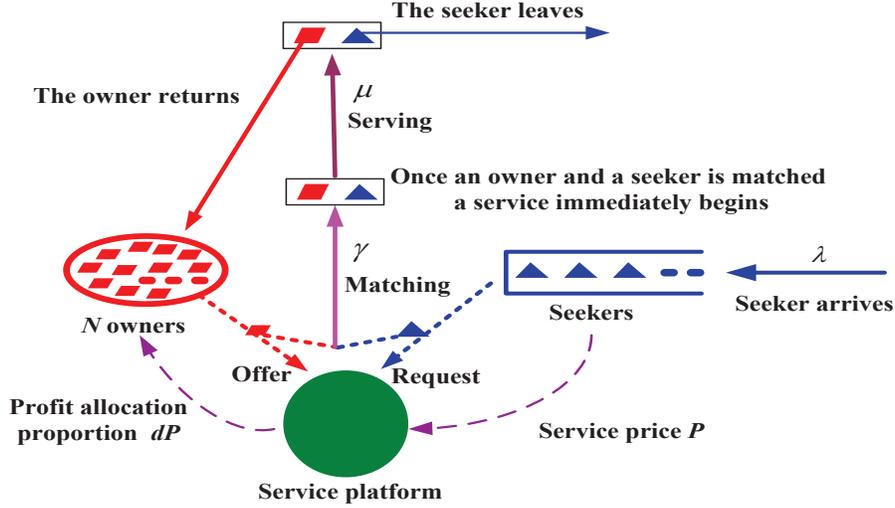}
\caption{The queueing structure of service platforms}%
\end{figure}

\begin{Rem}
Our queueing models have broad applications in the sharing economy (or the
service platform). In a bike-sharing system, the owners are bikes, and the
seekers are riders. In a car-sharing system, the owners are cars, and the
seekers are drivers. In Didi Taxi or Uber, the owners are taxis, and the
seekers are passengers. In a house-sharing system, the owners are houses or
rooms, and the seekers are tenants. In an equipment-sharing system, the owners
are equipments, and the seekers are customers.
\end{Rem}

\section{A QBD Process and Stability}

In this section, we express the first queueing model of service platforms as a
level-independent QBD process, and obtain a necessary and sufficient condition
under which this system is stable.

We denote by $N_{1}\left(  t\right)  $ and $N_{2}\left(  t\right)  $ the
number of seekers waiting for their services, and the number of idle owners
retained in the service platform at time $t\geq0$, respectively. Note that the
information completed in a service platform is observed at the matching
completion time (it is also the service beginning time), the first queueing
model of service platforms is modeled as a continuous-time Markov process
$\left\{  \left(  N_{1}\left(  t\right)  ,N_{2}\left(  t\right)  \right)
,\text{ }t\geq0\right\}  $ whose state space is given by%
\[
\Omega=\bigcup\limits_{k=0}^{\infty}S_{k},
\]
where the $0$th level is%
\[
S_{0}=\bigcup\limits_{i=0}^{N-1}\left\{  \left(  i,0\right)  ,\left(
i,1\right)  ,\left(  i,2\right)  ,...,\left(  i,N\right)  \right\}  ,
\]
and for $k\geq1$, the $k$th level is
\[
S_{k}=\left\{  \left(  N+k-1,0\right)  ,\left(  N+k-1,1\right)  ,...,\left(
N+k-1,N\right)  \right\}  .
\]
Since the information completed in a service platform is observed at the
matching completion time (it is also the service beginning time), we can
understand such special state transitions:%
\begin{align*}
&  \left(  k,k\right)  \text{ }\underrightarrow{k\gamma}\text{ }\left(
k-1,k-1\right)  ,\text{ \ }k\geq1,\\
&  \left(  k,k+j\right)  \text{ }\underrightarrow{k\gamma}\text{ }\left(
k-1,k+j-1\right)  ,\text{ \ }k\geq1,j\geq1,\\
&  \left(  k+j,k\right)  \text{ }\underrightarrow{k\gamma}\text{ }\left(
k+j-1,k-1\right)  ,\text{ \ }k\geq1,j\geq1.
\end{align*}
Based on this finding, the state transition relations of the Markov process
$\left\{  \left(  N_{1}\left(  t\right)  ,N_{2}\left(  t\right)  \right)
,\text{ }t\geq0\right\}  $ are depicted in Figure 2.

\begin{figure}[tbh]
\centering    \includegraphics[height=7cm
,width=14cm]{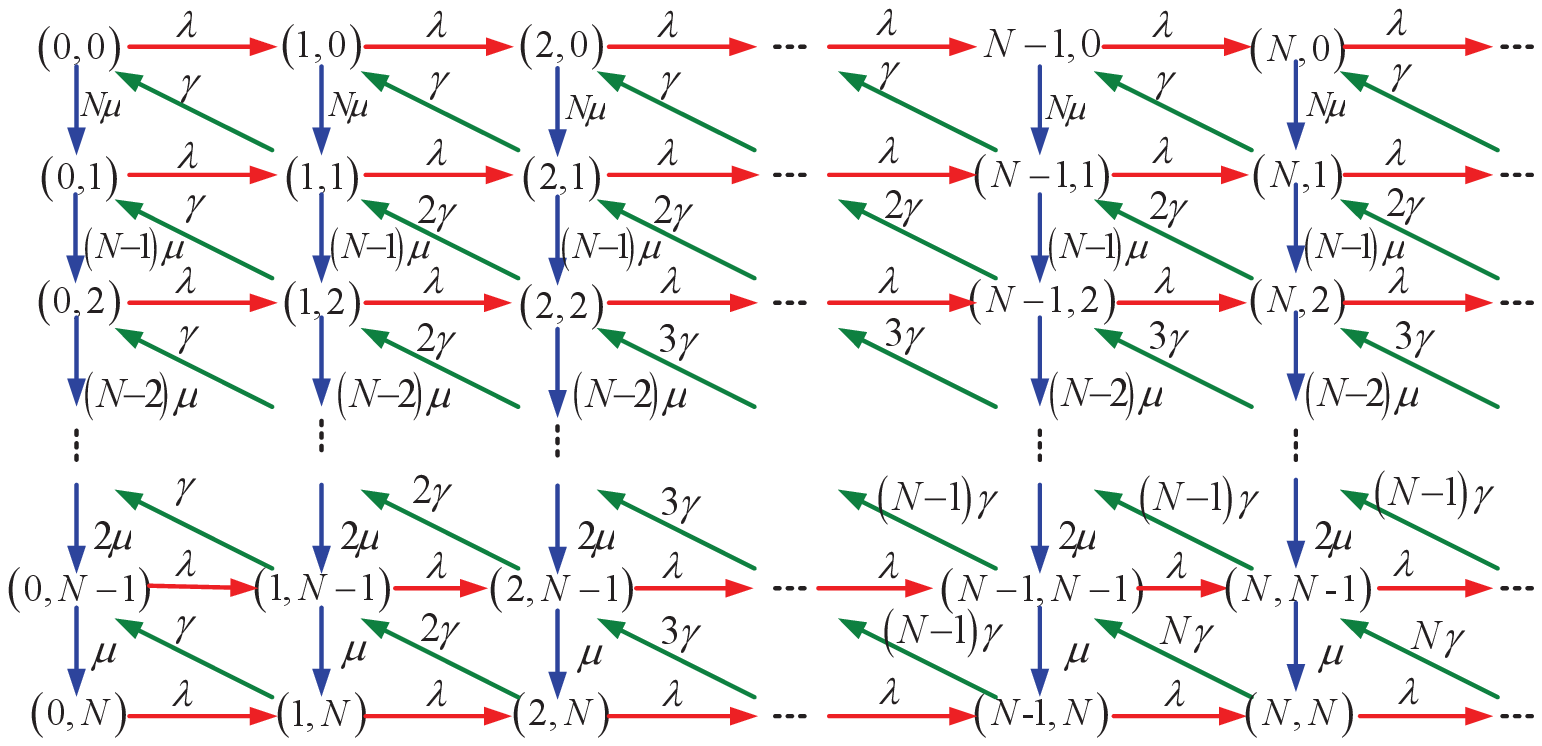}
\caption{The state transition relations of the Markov process in Model one}%
\end{figure}

It can be seen from Figure 2 that the infinitesimal generator of the Markov
process $\{(N_{1}\left(  t\right)  ,N_{2}\left(  t\right)  ),$ $t\geq0\}$ is
given by
\begin{equation}
Q=\left(
\begin{array}
[c]{cccccc}%
B_{0} & C_{0} &  &  &  & \\
A_{0} & B & C &  &  & \\
& A & B & C &  & \\
&  & A & B & C & \\
&  &  & \ddots & \ddots & \ddots
\end{array}
\right)  , \label{Q}%
\end{equation}
where the repeated blocks are given by%
\[
A=\left(
\begin{array}
[c]{cccccc}%
0 &  &  &  &  & \\
\gamma & 0 &  &  &  & \\
& 2\gamma & 0 &  &  & \\
&  & \ddots & \ddots &  & \\
&  &  & \left(  N-1\right)  \gamma & 0 & \\
&  &  &  & N\gamma & 0
\end{array}
\right)  ,\text{ }C=\left(
\begin{array}
[c]{cccccc}%
\lambda &  &  &  &  & \\
& \lambda &  &  &  & \\
&  & \lambda &  &  & \\
&  &  & \ddots &  & \\
&  &  &  & \lambda & \\
&  &  &  &  & \lambda
\end{array}
\right)  ,
\]%
\[
B=\left(
\begin{array}
[c]{cccccc}%
b^{\left(  0\right)  } & N\mu &  &  &  & \\
& b^{\left(  1\right)  } & \left(  N-1\right)  \mu &  &  & \\
&  & b^{\left(  2\right)  } & \left(  N-2\right)  \mu &  & \\
&  &  & \ddots & \ddots & \\
&  &  &  & b^{\left(  N-1\right)  } & \mu\\
&  &  &  &  & b^{\left(  N\right)  }%
\end{array}
\right)  ;
\]%
\[
b^{\left(  i\right)  }=-\left[  \lambda+\left(  N-i\right)  \mu+i\gamma
\right]  ,\text{ \ \ }0\leq i\leq N;
\]
and the boundary blocks are given by%
\[
A_{0}=\left(
\begin{array}
[c]{cccc}%
0 & \cdots & 0 & B_{2}^{\left(  N\right)  }%
\end{array}
\right)  ,\text{ \ }C_{0}=\left(
\begin{array}
[c]{c}%
0\\
\vdots\\
0\\
B_{0}^{\left(  N-1\right)  }%
\end{array}
\right)  ,
\]%
\[
B_{0}=\left(
\begin{array}
[c]{cccccc}%
B_{1}^{\left(  0\right)  } & B_{0}^{\left(  0\right)  } &  &  &  & \\
B_{2}^{\left(  1\right)  } & B_{1}^{\left(  1\right)  } & B_{0}^{\left(
1\right)  } &  &  & \\
& B_{2}^{\left(  2\right)  } & B_{1}^{\left(  2\right)  } & B_{0}^{\left(
2\right)  } &  & \\
&  & \ddots & \ddots & \ddots & \\
&  &  & B_{2}^{\left(  N-2\right)  } & B_{1}^{\left(  N-2\right)  } &
B_{0}^{\left(  N-2\right)  }\\
&  &  &  & B_{2}^{\left(  N-1\right)  } & B_{1}^{\left(  N-1\right)  }%
\end{array}
\right)  ,
\]%
\[
B_{0}^{\left(  k\right)  }=\left(
\begin{array}
[c]{ccc}%
\lambda &  & \\
& \ddots & \\
&  & \lambda
\end{array}
\right)  ,\text{ \ }0\leq k\leq N-1,\text{\ }%
\]%
\[
B_{2}^{\left(  k\right)  }=\left(
\begin{array}
[c]{cccccccc}%
0 &  &  &  &  &  &  & \\
\gamma & 0 &  &  &  &  &  & \\
& 2\gamma & 0 &  &  &  &  & \\
&  & \ddots & \ddots &  &  &  & \\
&  &  & \left(  k-1\right)  \gamma & 0 &  &  & \\
&  &  &  & k\gamma & 0 &  & \\
&  &  &  &  & \ddots & 0 & \\
&  &  &  &  &  & k\gamma & 0
\end{array}
\right)  ,\text{ \ }1\leq k\leq N,
\]
and for $0\leq k\leq N-1$%
\[
B_{1}^{\left(  k\right)  }=\left(
\begin{array}
[c]{cccccccc}%
c^{\left(  0\right)  } & N\mu &  &  &  &  &  & \\
& c^{\left(  1\right)  } & \left(  N-1\right)  \mu &  &  &  &  & \\
&  & \ddots & \ddots &  &  &  & \\
&  &  & c^{\left(  k-1\right)  } & \left(  N-k+1\right)  \mu &  &  & \\
&  &  &  & c^{\left(  k\right)  } & \left(  N-k\right)  \mu &  & \\
&  &  &  &  & \ddots & \ddots & \\
&  &  &  &  &  & c^{\left(  N-1\right)  } & \mu\\
&  &  &  &  &  &  & c^{\left(  N\right)  }%
\end{array}
\right)  ,
\]%
\[
c^{\left(  i\right)  }=\left\{
\begin{array}
[c]{ll}%
-\left[  \lambda+i\gamma+\left(  N-i\right)  \mu)\right]  , & 0\leq i\leq
k-1,\\
-\left[  \lambda+k\gamma+\left(  N-i\right)  \mu)\right]  , & k\leq i\leq N.
\end{array}
\right.
\]

\vskip                                                  0.4cm

It is easy to see from the infinitesimal generator $Q$ that the Markov process
$\{(N_{1}\left(  t\right)  ,N_{2}\left(  t\right)  ),$ $t\geq0\}$ is an
irreducible QBD process.

To find the stable condition of the QBD process $Q$, the following lemma is
useful in our later computation. The first equation in Lemma 1 is
straightforward by means of the Newton binomial theorem, and the second
equation can be obtained by taking the derivatives in both sides of the first equation.

\begin{Lem}
For $x>0$, we have%
\[
\sum\limits_{i=0}^{N}C_{N}^{i}x^{i}=\left(  1+x\right)  ^{N}%
\]
and%
\[
\sum\limits_{i=1}^{N}iC_{N}^{i}x^{i}=Nx\left(  1+x\right)  ^{N-1}.
\]
\end{Lem}

The following theorem provides a necessary and sufficient condition under
which the QBD process $Q$ is stable.

\begin{The}
\ \label{The1}The QBD process $Q$ is positive recurrent if and only if
$\rho=\lambda\left(  1+\mu/\gamma\right)  /N\mu<1$. Also, the first queueing
model of the service platforms is stable.
\end{The}

\noindent\textbf{Proof.}\textit{ }It is easy to identify the irreducibility of
the QBD process $Q$ from Figure 2.

To prove that the QBD process $Q$ is positive recurrent, it is the key to find
a necessary and sufficient condition by using the mean drift technique by
Neuts \cite{Neu:1981}.

Let $D=A+B+C$. Then%
\[
D=\left(
\begin{array}
[c]{cccccc}%
d^{\left(  0\right)  } & N\mu &  &  &  & \\
\gamma & d^{\left(  1\right)  } & \left(  N-1\right)  \mu &  &  & \\
& 2\gamma & d^{\left(  2\right)  } & \left(  N-2\right)  \mu &  & \\
&  & \ddots & \ddots & \ddots & \\
&  &  & \left(  N-1\right)  \gamma & d^{\left(  N-1\right)  } & \mu\\
&  &  &  & N\gamma & d^{\left(  N\right)  }%
\end{array}
\right)  ,
\]
where%
\[
d^{\left(  i\right)  }=-\left[  \left(  N-i\right)  \mu+i\gamma\right]
,\text{ \ }0\leq i\leq N.
\]
Note that the Markov process $D$ is irreducible and has finite states. Thus,
it is positive recurrent. Let $\alpha=\left(  \alpha_{1},\alpha_{2},\alpha
_{3},\ldots,\alpha_{N+1}\right)  $ be the stationary probability vector of the
Markov process $D$. Then
\[
\alpha D=\mathbf{0},\text{ \ }\alpha\mathbf{e}=1,
\]
where $\mathbf{0}$ is a zero row vector with a suitable sizes (here, size
$N+1)$, and $\mathbf{e}$ is a column vector of ones with a suitable sizes
(here, size $N+1)$. Note that $\alpha>0$ is due to the fact that the Markov
process $D$ is irreducible.

Once the stationary probability vector $\alpha$ is obtained, we can compute
the (upward and downward) mean drift rates of the QBD process $Q$. From level
$i$ to level $i+1$, the upward mean drift rate is given by%
\[
\alpha C\mathbf{e=}\lambda\alpha\mathbf{e}=\lambda,
\]
since $C=\lambda I$ and $\alpha\mathbf{e}=1$, where $I$ is an identity matrix.

Similarly, from level $i$ to level $i-1$, the downward mean drift rate is
$\alpha A\mathbf{e}$. To compute the drift rate $\alpha A\mathbf{e}$, we need
to solve the system of linear equations: $\alpha D=\mathbf{0}$ and
$\alpha\mathbf{e}=1$ as follows:%
\[
\alpha_{1}=\frac{1}{\sum\limits_{i=0}^{N}C_{N}^{i}\left(  \frac{\mu}{\gamma
}\right)  ^{i}},
\]
and for $2\leq k\leq N+1$,%
\[
\alpha_{k}=C_{N}^{k-1}\left(  \frac{\mu}{\gamma}\right)  ^{k-1}\alpha
_{1}=\frac{C_{N}^{k-1}\left(  \frac{\mu}{\gamma}\right)  ^{k-1}}%
{\sum\limits_{i=0}^{N}C_{N}^{i}\left(  \frac{\mu}{\gamma}\right)  ^{i}}.
\]
Now, we compute%
\begin{align*}
\alpha A\mathbf{e}  &  =1\gamma\alpha_{2}+2\gamma\alpha_{3}+\cdots
+N\gamma\alpha_{N+1}\\
&  =\frac{\gamma\sum\limits_{i=1}^{N}iC_{N}^{i}\left(  \frac{\mu}{\gamma
}\right)  ^{i}}{\sum\limits_{i=0}^{N}C_{N}^{i}\left(  \frac{\mu}{\gamma
}\right)  ^{i}}=\frac{N\mu\gamma}{\mu+\gamma}.
\end{align*}
Let $\rho=\left(  \alpha C\mathbf{e}\right)  /\left(  \alpha A\mathbf{e}%
\right)  $. Then
\[
\rho=\frac{\lambda\left(  \mu+\gamma\right)  }{N\mu\gamma}.
\]
It is clear that $\alpha C\mathbf{e<}$ $\alpha A\mathbf{e}$ if and only if
$\rho<1$.\textbf{ }Therefore, the QBD process $Q$ is positive recurrent if and
only if $\rho<1$ or $\alpha C\mathbf{e<}$ $\alpha A\mathbf{e}$. This completes
the proof. \hfill$\blacksquare$

The following corollary provides a novel stable condition under which the QBD
process $Q$ is positive recurrent. We show that the first queueing model of
service platforms is always stable as long as the number $N$ of independent
owners is large enough. Such a result is not intuitive but it is very useful
in the design of a service platform whose normal operation needs to have
enough independent owners.

\begin{Cor}
If $N>1+\left\lfloor \frac{\lambda}{\mu}\left(  1+\frac{\mu}{\gamma}\right)
\right\rfloor $, where $\left\lfloor x\right\rfloor $ is the maximum integer
lower than or equal to $x$, then the QBD process $Q$ (or the first queueing
model of service platforms) must be positive recurrent.
\end{Cor}

\noindent\textbf{Proof.}\textit{ }We only need to observe that if
$N>1+\left\lfloor \frac{\lambda}{\mu}\left(  1+\frac{\mu}{\gamma}\right)
\right\rfloor $, then $\rho<1$. This completes the proof. \hfill$\blacksquare$

\section{Performance Measures}

In this section, we apply the matrix-geometric solution to give the stationary
probability vector of the QBD process, and provide some useful performance
measures of the service platform.

We write%
\[
p_{i,j}\left(  t\right)  =P\left\{  N_{1}\left(  t\right)  =i,N_{2}\left(
t\right)  =j\right\}  .
\]
Since the QBD process is stable, we have%
\[
\pi_{i,j}=\lim_{t\rightarrow+\infty}p_{i,j}\left(  t\right)  .
\]
For $k=0$, we write%
\[
\pi_{0}=\left(  \pi_{0}^{\left(  0\right)  },\pi_{0}^{\left(  1\right)  }%
,\pi_{0}^{\left(  2\right)  },\ldots,\pi_{0}^{\left(  N-1\right)  }\right)
\]
,%
\[
\pi_{0}^{\left(  i\right)  }=\left(  \pi_{i,0},\pi_{i,1},\ldots,\pi
_{i,N}\right)  ,\text{ \ }0\leq i\leq N-1;
\]
for $k\geq1$, we write%
\[
\pi_{k}=\left(  \pi_{N+k-1,0},\pi_{N+k-1,1},\pi_{N+k-1,2},\ldots,\pi
_{N+k-1,N}\right)  ,
\]
and%
\[
\pi=\left(  \pi_{0},\pi_{1},\pi_{2},\ldots\right)  .
\]
If $\pi=\left(  \pi_{0},\pi_{1},\pi_{2},\ldots\right)  $ is the stationary
probability vector of the QBD process $Q$, then $\pi$ satisfies the system of
linear equations:%
\begin{equation}
\pi Q=\mathbf{0,}\text{ \ }\pi\mathbf{e}=1, \label{equ1}%
\end{equation}
which leads to%
\begin{equation}
\pi_{0}B_{0}+\pi_{1}A_{0}=0, \label{equ2}%
\end{equation}%
\begin{equation}
\pi_{0}C_{0}+\pi_{1}B+\pi_{2}A=0, \label{equ3}%
\end{equation}%
\begin{equation}
\pi_{k-1}C+\pi_{k}B+\pi_{k+1}A=0,\text{ \ }k\geq2, \label{equ4}%
\end{equation}%
\begin{equation}
\sum\limits_{k=0}^{\infty}\pi_{k}\mathbf{e}=1. \label{equ5}%
\end{equation}

Let the matrix rate $R$ be the minimal non-negative solution to the matrix
quadratic equation%
\begin{equation}
R^{2}A+RB+C=0. \label{equ7}%
\end{equation}

The following theorem can directly be obtained by using Chapter 3 of Neuts
\cite{Neu:1981}.

\begin{The}
The stationary probability vector of the QBD process $Q$ is a matrix-geometric
solution
\begin{equation}
\pi_{i}=\pi_{1}R^{i-1},\text{ }i\geq1, \label{equ6}%
\end{equation}
where $\left(  \pi_{0},\pi_{1}\right)  $ is the unique solution to the
following system of linear equations%
\[
\pi_{0}B_{0}+\pi_{1}A_{0}=0,
\]%
\[
\pi_{0}C_{0}+\pi_{1}\left(  B+RA\right)  =0
\]
and%
\[
\pi_{0}e+\pi_{1}\left(  I-R\right)  ^{-1}\mathbf{e}=1.
\]
\end{The}

\subsection{The stationary performance measures}

In this subsection, by using the stationary probability vector of the QBD
process $Q$, we provide some useful performance measures of the first queueing
model of service platforms, e.g., the stationary average numbers, and the
expected profits of the platform and each owner.

\textbf{(a) The stationary average numbers}

Let $\mathcal{Q}^{\left(  1\right)  }$ and $\mathcal{Q}^{\left(  2\right)  }$
denote the stationary numbers of the idle owners retained in the service
platform, and of the seekers waiting for their services, respectively. Then%
\begin{align}
E\left[  \mathcal{Q}^{\left(  1\right)  }\right]    & =\sum\limits_{i=0}%
^{\infty}\sum\limits_{j=1}^{N}j\pi_{i,j}\nonumber\\
& =\sum_{l=0}^{N-1}\pi_{0}^{\left(  l\right)  }f+\pi_{1}\left(  I-R\right)
^{-1}f,\label{equ10}%
\end{align}
where $f=\left(  0,1,2,\ldots,N\right)  ^{T}$, and%
\begin{align}
E\left[  \mathcal{Q}^{\left(  2\right)  }\right]    & =\sum\limits_{i=1}%
^{\infty}\sum\limits_{j=0}^{N}i\pi_{i,j}\nonumber\\
& =\pi_{1}\left(  I-R\right)  ^{-1}\mathbf{e}.\label{equ9}%
\end{align}

\textbf{(b) The expected profits of the service platform and each owner}

Let $f_{1}$ and $f_{2}$ denote the expected profits per unit time of the
service platform and each owner, respectively. Then by using the service price
$P$ and the profit allocation proportion $d$, we have%
\[
f_{1}=\left(  1-d\right)  \cdot P\left(  N-E\left[  \mathcal{Q}^{\left(
1\right)  }\right]  \right)  \mu,
\]
and%
\begin{align*}
f_{2}  &  =\frac{1}{N}d\cdot P\left(  N-E\left[  \mathcal{Q}^{\left(
1\right)  }\right]  \right)  \mu\\
&  =d\cdot P\left(  1-\frac{1}{N}E\left[  \mathcal{Q}^{\left(  1\right)
}\right]  \right)  \mu,
\end{align*}
where $E\left[  \mathcal{Q}^{\left(  1\right)  }\right]  $ is given in
(\ref{equ10}). Note that $N-E\left[  \mathcal{Q}^{\left(  1\right)  }\right]
$ is the average number of the busy owners, and $P\left(  N-E\left[
\mathcal{Q}^{\left(  1\right)  }\right]  \right)  \mu$ is the total expected
profits per unit time of the service platform, which is paid by all the
seekers served per unit time.

\subsection{The sojourn time of an arriving seeker}

In this subsection, we compute the expected sojourn time of each arriving
seeker at the service platform. That is, the sojourn time begins from the
moment of the seeker entering the service platform to the epoch that the
seeker leaves the system after receiving his/her service from the owner.

We denote by $W$ the sojourn time of an arriving seeker at the service
platform. To compute the expected sojourn time $E\left[  W\right]  $, we need
to apply the first passage time of the QBD process with an absorbing state. To
this end, we revise the QBD process $\left\{  \left(  N_{1}\left(  t\right)
,N_{2}\left(  t\right)  \right)  ,\text{ }t\geq0\right\}  $ as a QBD process
with an absorbing state
\[
\Delta=\left\{  \left(  0,0\right)  ,\left(  0,1\right)  ,\left(  0,2\right)
,...,\left(  0,N-1\right)  \right\}  ,
\]
while state $\left(  0,N\right)  $ is a sliding state and is mitted here.
Thus, the QBD process with the absorption state $\Delta$ is depicted in Figure
3. Based on this, the state space of the QBD process with the absorbing state
is given by%
\[
\widetilde{\Omega}=\left\{  \Delta\right\}  \cup\widetilde{S}_{0}\cup\left\{
\bigcup\limits_{i=1}^{\infty}S_{i}\right\}  ,
\]
where%
\[
\widetilde{S}_{0}=\bigcup\limits_{i=1}^{N-1}\left\{  \left(  i,0\right)
,\left(  i,1\right)  ,\left(  i,2\right)  ,...,\left(  i,N\right)  \right\}
,
\]
and for $k\geq1$,
\[
S_{k}=\left\{  \left(  N+k-1,0\right)  ,\left(  N+k-1,1\right)  ,...,\left(
N+k-1,N\right)  \right\}  .
\]

\begin{figure}[tbh]
\centering           \includegraphics[height=7cm
,width=14cm]{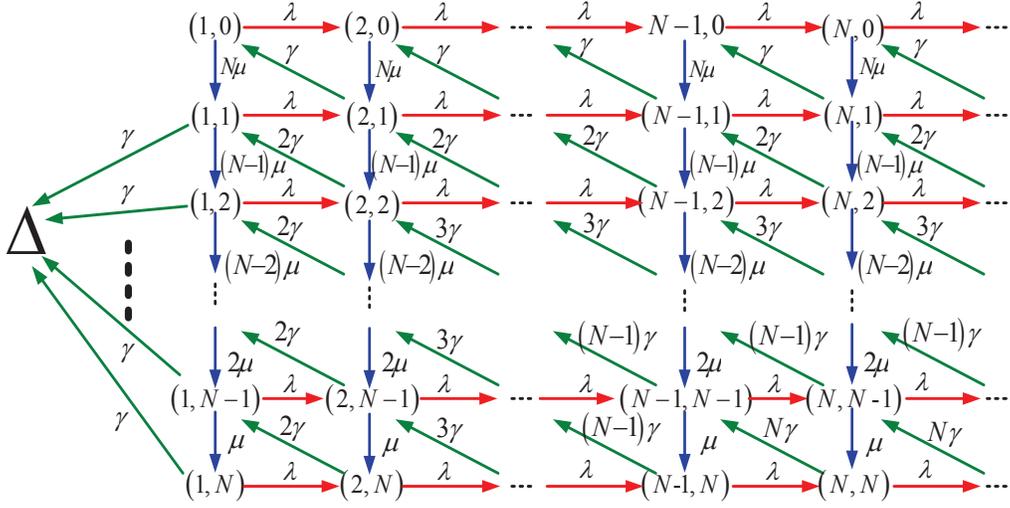}
\caption{The state transition relations of the QBD process with an absorbing
state}%
\end{figure}

From Figure 3, it is seen that the infinitesimal generator of the QBD process
with the absorption state $\Delta$ is given by%
\begin{equation}
\mathbf{Q}=\left(
\begin{array}
[c]{cccccc}%
0 & \mathbf{0} & \mathbf{0} & \mathbf{0} & \mathbf{0} & \cdots\\
\widetilde{\varphi} & \widetilde{B}_{0} & \widetilde{C}_{0} &  &  & \\
& \widetilde{A}_{0} & B & C &  & \\
&  & A & B & C & \\
&  &  & \ddots & \ddots & \ddots
\end{array}
\right)  \label{equ11}%
\end{equation}
where the blocks $A$, $B$ and $C$ are the same as those given in matrix $Q$,
and%
\[
\widetilde{B}_{0}=\left(
\begin{array}
[c]{cccccc}%
B_{1}^{\left(  1\right)  } & B_{0}^{\left(  1\right)  } &  &  &  & \\
B_{2}^{\left(  2\right)  } & B_{1}^{\left(  2\right)  } & B_{0}^{\left(
2\right)  } &  &  & \\
& \ddots & \ddots & \ddots &  & \\
&  &  & B_{2}^{\left(  N-2\right)  } & B_{1}^{\left(  N-2\right)  } &
B_{0}^{\left(  N-2\right)  }\\
&  &  &  & B_{2}^{\left(  N-1\right)  } & B_{1}^{\left(  N-1\right)  }%
\end{array}
\right)  ,
\]%
\[
\widetilde{A}_{0}=\left(
\begin{array}
[c]{cccc}%
0 & \cdots & 0 & B_{2}^{\left(  N\right)  }%
\end{array}
\right)  ,\text{ \ }\widetilde{C}_{0}=\left(
\begin{array}
[c]{c}%
0\\
\vdots\\
0\\
B_{0}^{\left(  N-1\right)  }%
\end{array}
\right)  ,
\]%
\[
\widetilde{\varphi}=\left(
\begin{array}
[c]{c}%
\varphi\\
0\\
\vdots\\
0
\end{array}
\right)  ,\text{ \ }\varphi=\left(
\begin{array}
[c]{c}%
0\\
\gamma\\
\vdots\\
\gamma
\end{array}
\right)  .
\]

Let%
\[
T=\left(
\begin{array}
[c]{ccccc}%
\widetilde{B}_{0} & \widetilde{C}_{0} &  &  & \\
\widetilde{A}_{0} & B & C &  & \\
& A & B & C & \\
&  & \ddots & \ddots & \ddots
\end{array}
\right)  ,\text{ \ }T^{0}=\left(
\begin{array}
[c]{c}%
\widetilde{\varphi}\\
0\\
0\\
\vdots
\end{array}
\right)  .
\]
Then the infinitesimal generator $\mathbf{Q}$ can be abbreviated as%
\begin{equation}
\mathbf{Q}=\left(
\begin{array}
[c]{cc}%
0 & \mathbf{0}\\
T^{0} & T
\end{array}
\right)  . \label{equ12}%
\end{equation}
It is easy to see that $T\mathbf{e}+T^{0}=0$.

Now, the initial probability vector of the QBD process $\mathbf{Q}$ with the
absorption state $\Delta$ is written as%
\[
\omega=\left(  \omega_{\triangle},\widetilde{\omega}\right)  ,
\]
where $\omega_{\triangle}$ is a scalar, and $\omega_{0}^{\left(  i\right)  }$
for $1\leq i\leq N-1$ and $\omega_{k}$ for $k\geq1$ are row vectors of size
$N+1$,%
\[
\widetilde{\omega}=\left(  \omega_{0},\omega_{1},\omega_{2},\ldots\right)  ,
\]%
\[
\omega_{0}=\left(  \omega_{0}^{\left(  1\right)  },\omega_{0}^{\left(
2\right)  },\omega_{0}^{\left(  3\right)  },\ldots,\omega_{0}^{\left(
N-1\right)  }\right)  ,
\]
with%
\[
\omega_{\triangle}+\sum_{i=1}^{N-1}\omega_{0}^{\left(  i\right)  }%
\mathbf{e}+\sum_{k=1}^{\infty}\omega_{k}\mathbf{e}=1,
\]

By using the stationary probability vector $\pi=\left(  \pi_{0},\pi_{1}%
,\pi_{2},\ldots\right)  $, we can set up a key initial probability vector%
\[
\omega=\left(  \omega_{\triangle};\omega_{0}^{\left(  1\right)  },\omega
_{0}^{\left(  2\right)  },\omega_{0}^{\left(  3\right)  },\ldots,\omega
_{0}^{\left(  N-1\right)  };\omega_{1},\omega_{2},\ldots\right)  ,
\]
where
\begin{align*}
\omega_{\triangle} &  =\pi_{0}^{\left(  0\right)  }\mathbf{e,}\\
\omega_{0}^{\left(  i\right)  } &  =\pi_{0}^{\left(  i\right)  },\text{ }1\leq
i\leq N-1,\\
\omega_{k} &  =\pi_{k},\text{ \ }k\geq1.
\end{align*}

The following theorem uses the phase-type distribution of infinite size to
provide expression for the probability distribution of the sojourn time $W$.

\begin{The}
\label{The:PH} If the initial probability vector of the QBD process
$\mathbf{Q}$ with the absorption state $\Delta$ is $\omega=\left(
\omega_{\triangle},\widetilde{\omega}\right)  $, then

(1) the probability distribution of the sojourn time $W$ is of phase type with
an irreducible representation $\left(  \widetilde{\omega},T\right)  $, i.e.,
\[
F_{W}\left(  t\right)  =P\left\{  W\leq t\right\}  =1-\omega_{\triangle
}-\widetilde{\omega}\exp\left\{  Tt\right\}  \mathbf{e,}\text{ \ }t\geq0;
\]

(2) the expected sojourn time%
\[
E\left[  W\right]  =\widetilde{\omega}\left(  -T\right)  _{\min}%
^{-1}\mathbf{e,}%
\]
where $\left(  -T\right)  _{\min}^{-1}$ is the minimal nonnegative inverse
matrix of the matrix $-T$.
\end{The}

\noindent\textbf{Proof.}\textit{ }(1) For $i=1,2,3,...$, and $j=1,2,3,...,N$,
we write%
\[
q_{i,j}\left(  t\right)  =P\left\{  N_{1}\left(  t\right)  =i,N_{2}\left(
t\right)  =j\right\}  ,
\]
For $k=0$, we write%
\[
q_{0}\left(  t\right)  =\left(  q_{0}^{\left(  1\right)  }\left(  t\right)
,q_{0}^{\left(  2\right)  }\left(  t\right)  ,q_{0}^{\left(  3\right)
}\left(  t\right)  ,\ldots,q_{0}^{\left(  N-1\right)  }\left(  t\right)
\right)
\]
,%
\[
q_{0}^{\left(  i\right)  }\left(  t\right)  =\left(  q_{i,1}\left(  t\right)
,q_{i,2}\left(  t\right)  ,q_{i,3}\left(  t\right)  ,\ldots,q_{i,N}\left(
t\right)  \right)  ,\text{ \ }1\leq i\leq N-1;
\]
for $k\geq1$, we write%
\[
q_{k}\left(  t\right)  =\left(  q_{N+k-1,0}\left(  t\right)  ,q_{N+k-1,1}%
\left(  t\right)  ,q_{N+k-1,2}\left(  t\right)  ,\ldots,q_{N+k-1,N}\left(
t\right)  \right)  ,
\]
and%
\[
q\left(  t\right)  =\left(  q_{0}\left(  t\right)  ,q_{1}\left(  t\right)
,q_{2}\left(  t\right)  ,\ldots\right)  .
\]
It follows from the Chapman-Kolmogorov forward equation that
\begin{equation}
\frac{d}{dt}q\left(  t\right)  =q\left(  t\right)  T,\label{equ20}%
\end{equation}
with the initial condition%
\begin{equation}
q\left(  0\right)  =\widetilde{\omega}.\label{equ21}%
\end{equation}
It follows from (\ref{equ20}) and (\ref{equ21}) that%
\begin{equation}
q\left(  t\right)  =\widetilde{\omega}e^{Tt}.\label{equ22}%
\end{equation}
Note that $q\left(  0\right)  \mathbf{e}=1-\omega_{\triangle}$. We then
obtain\qquad%
\begin{align*}
F_{W}\left(  t\right)   &  =P\left\{  W\leq t\right\}  =1-\omega_{\triangle
}-q\left(  t\right)  \mathbf{e}\\
&  =1-\omega_{\triangle}-\widetilde{\omega}\exp\left(  Tt\right)
\mathbf{e,}\text{ }t\geq0.
\end{align*}

(2) Now, we compute the expected sojourn time $E\left[  W\right]  $ by the
Laplace-Stieltjes transform. Let $f\left(  s\right)  $ be the
Laplace-Stieltjes transform of the distribution function $F_{W}\left(
t\right)  $ or the random variable $W$. Then%
\begin{align}
f\left(  s\right)   &  =\int\nolimits_{0}^{\infty}e^{-st}dF_{W}(t)\nonumber\\
&  =1-\omega_{\triangle}+\widetilde{\omega}\left(  sI-T\right)  _{\min}%
^{-1}T^{0},\label{equf1}%
\end{align}
where $\left(  sI-T\right)  _{\min}^{-1}$ is the minimal nonnegative inverse
of the matrix $sI-T$ of infinite size. It is easy to see that
\begin{equation}
E\left[  W\right]  =-\frac{d}{ds}f\left(  s\right)  _{\mid s=0}=\widetilde
{\omega}\left[  \left(  sI-T\right)  _{\min}^{-2}\right]  _{\mid s=0}%
T^{0}=\widetilde{\omega}\left(  -T\right)  _{\min}^{-1}\mathbf{e,}%
\label{equf2}%
\end{equation}
by using $T\mathbf{e}+T^{0}=0$ and $\left(  -T\right)  _{\min}^{-1}\left(
-T\right)  \mathbf{e=e}$. This completes the proof. \hfill$\blacksquare$

In the remainder of this section, we use the RG-factorizations by Li
\cite{Li:2010} to compute the expected sojourn time $E\left[  W\right]  $. It
is easy to see that the key is how to deal with the minimal nonnegative
inverse matrix $\left(  -T\right)  _{\min}^{-1}$ of the matrix $-T$ of
infinite size. To this end, we write%
\[
T=\left(
\begin{array}
[c]{c}%
T_{1,1}\text{ }T_{1,2}\\
T_{2,1}\text{ }T_{2,2}%
\end{array}
\right)  ,
\]
where%
\[
T_{1,1}=\widetilde{B}_{0},\text{ \ }T_{1,2}=\left(
\begin{array}
[c]{ccccc}%
\widetilde{C}_{0}\text{ } &  &  &  &
\end{array}
\text{\ }\right)  ,\text{ \ }T_{2,1}=\widetilde{A}_{0},
\]%
\[
T_{2,2}=\left(
\begin{array}
[c]{ccccc}%
B & C &  &  & \\
A & B & C &  & \\
& A & B & C & \\
&  & \ddots & \ddots & \ddots
\end{array}
\right)  .
\]

Note that the QBD process $Q$ is irreducible, and thus matrix $T$ and
$T_{2,2}$\ must be invertible. Here, we take that $T^{-1}=T_{\max}^{-1}$ and
$T_{2,2}^{-1}=\left(  T_{2,2}^{-1}\right)  _{\max}$, i.e., their inverse
matrices are maximal non-positive. Based on this, it is easy to check that%
\begin{equation}
T^{-1}=\left[
\begin{array}
[c]{cc}%
T_{1,1;2}^{-1} & -T_{1,1;2}^{-1}T_{1,2}T_{2,2}^{-1}\\
-T_{2,2}^{-1}T_{2,1}T_{1,1;2}^{-1} & T_{2,2}^{-1}+T_{2,2}^{-1}T_{2,1}%
T_{1,1;2}^{-1}T_{1,2}T_{2,2}^{-1}%
\end{array}
\right]  , \label{T1}%
\end{equation}
where%
\begin{equation}
T_{1,1;2}=T_{1,1}-T_{1,2}T_{2,2}^{-1}T_{2,1}. \label{T2}%
\end{equation}
It is clear that the inverse matrix $T^{-1}$ can be expressed by means of the
inverse matrix $T_{2,2}^{-1}$ of matrix $T_{2,2}$. This relation plays a key
role in setting up the PH distribution of infinite sizes with an irreducible
representation $\left(  \widetilde{\omega},T\right)  $.

To compute the inverse matrix $T_{2,2}^{-1}$ of infinite sizes, we define the
UL-type $U$-, $R$- and $G$-measures as follows. Let $R$ and $G$ be the minimal
non-negative solutions to the matrix quadratic equations%
\begin{equation}
R^{2}A+RB+C=0 \label{TR}%
\end{equation}
and%
\begin{equation}
A+BG+CG^{2}=0, \label{TG}%
\end{equation}
respectively. The $U$-measure is given by
\begin{equation}
U=B+RA=B+CG\text{.} \label{TU1}%
\end{equation}
Now, we can provide the UL-type RG-factorization of the Markov process
$T_{2,2}$ of infinite size as follows:%
\begin{equation}
T_{2,2}=\left(  I-R_{U}\right)  U_{D}\left(  I-G_{L}\right)  , \label{equ25}%
\end{equation}
where%
\begin{equation}
I-R_{U}=\left(
\begin{array}
[c]{ccccc}%
I & -R &  &  & \\
& I & -R &  & \\
&  & I & -R & \\
&  &  & I & \ddots\\
&  &  &  & \ddots
\end{array}
\right)  , \label{equ251}%
\end{equation}%
\begin{equation}
U_{D}=\text{diag}\left(  U,U,U,U,\ldots\right)  , \label{equ252}%
\end{equation}%
\begin{equation}
I-G_{L}=\left(
\begin{array}
[c]{ccccc}%
I &  &  &  & \\
-G & I &  &  & \\
& -G & I &  & \\
&  & -G & I & \\
&  &  & \ddots & \ddots
\end{array}
\right)  . \label{equ253}%
\end{equation}

It is easy to check that the matrices $I-R_{U}$, $U_{D}$, and $I-G_{L}$ are
invertible, and%
\begin{equation}
\left(  I-R_{U}\right)  ^{-1}=\left(
\begin{array}
[c]{ccccc}%
I & R & R^{2} & R^{3} & \ldots\\
& I & R & R^{2} & \ldots\\
&  & I & R & \ldots\\
&  &  & I & \ldots\\
&  &  &  & \ddots
\end{array}
\right)  , \label{equ254}%
\end{equation}%
\begin{equation}
U_{D}^{-1}=\text{diag}\left(  U^{-1},U^{-1},U^{-1},U^{-1},\ldots\right)  ,
\label{equ255}%
\end{equation}%
\begin{equation}
\left(  I-G_{L}\right)  ^{-1}=\left(
\begin{array}
[c]{ccccc}%
I &  &  &  & \\
G & I &  &  & \\
G^{2} & G & I &  & \\
G^{3} & G^{2} & G & I & \\
\vdots & \vdots & \vdots & \vdots & \ddots
\end{array}
\right)  . \label{equ256}%
\end{equation}
By using the RG-factorization (\ref{equ25}), we obtain%
\begin{equation}
T_{2,2}^{-1}=\left(  I-G_{L}\right)  ^{-1}U_{D}^{-1}\left(  I-R_{U}\right)
^{-1}. \label{T3}%
\end{equation}
By means of (\ref{T1}), (\ref{T2}), and (\ref{T3}), we can obtain the inverse
matrix $T^{-1}$.

\section{The Second Queueing Model}

In this section, we provide a simple analysis for the second queueing model of
service platforms by using the level-independent QBD process.

We first observe the matching and service processes. Let $X$ and $Y$ be two
exponential random variables with the means $1/\gamma$ and $1/\mu$,
respectively. Then the sum $X+Y$ follows a generalized Erlang distribution of
order $2$, or $X+Y$ also follows a phase-type (PH) distribution of order $2$
under a more general setting. Here, $X$ and $Y$ are regarded as Phases $1$ and
$2$ of the PH distribution, respectively. For the generalized Erlang
distribution of order $2$, we write%
\[
\alpha=\left(  1,0\right)  ,\text{ }T=\left(
\begin{array}
[c]{cc}%
-\gamma & \gamma\\
0 & -\mu
\end{array}
\right)  ,\text{ }T^{0}=\left(
\begin{array}
[c]{c}%
0\\
\mu
\end{array}
\right)  ,\text{ }\mathbf{e}=\left(
\begin{array}
[c]{c}%
1\\
1
\end{array}
\right)  .
\]
For $2\leq n\leq N$%
\[
T\left(  n\right)  =\underset{\text{Keronecker sum of }n\text{ matrices }%
T}{\underbrace{T\oplus T\oplus\cdots\oplus T};}%
\]
for $2\leq k\leq N-1$%
\begin{align*}
C\left(  k+1\right)  =  &  T^{0}\otimes\underset{\text{Keronecker product of
}k\text{ identity matrices}}{\underbrace{I\otimes I\otimes\cdots\otimes I}%
}+\mathbf{e}\otimes\left(  T^{0}\alpha\right)  \otimes I\otimes\cdots\otimes
I+\cdots\\
&  +\mathbf{e}\otimes I\otimes\cdots\otimes\left(  T^{0}\alpha\right)  \otimes
I+\mathbf{e}\otimes I\otimes I\otimes\cdots\otimes\left(  T^{0}\alpha\right)
,
\end{align*}
and for $1\leq l\leq N-1$%
\[
D\left(  l\right)  =\underset{\text{Keronecker product of }l\text{ identity
matrix}}{\underbrace{I\otimes I\otimes\cdots\otimes I}}\otimes\alpha,
\]

For the second queueing model of service platforms, the information completed
in a service platform is observed at the matching beginning time. Thus, the
sum $X+Y$ of the matching and service times is regarded as a generalized
service time, which follows a generalized Erlang distribution of order $2$
(abbreviated as $\widetilde{\text{E}}_{2}$). In this case, the second queueing
model of service platforms can be regarded as an M$/\widetilde{\text{E}}%
_{2}/N$ queue (or an M$/$PH$/N$ queue).

To study the M$/\widetilde{\text{E}}_{2}/N$ queue, we denote by $N\left(
t\right)  $ and $M\left(  t\right)  $ the number of seekers waiting for their
services, and the number of working owners at time $t\geq0$, respectively. If
$N\left(  t\right)  =0$, then $0\leq M\left(  t\right)  \leq N$; and if
$N\left(  t\right)  \geq1$, then $M\left(  t\right)  =N$. Let $J\left(
t\right)  $ be the phase of the generalized Erlang service time $\widetilde
{\text{E}}_{2}$ at time $t$. Then $J\left(  t\right)  \in\left\{  1,2\right\}
$. It is easy to see that $\left\{  N\left(  t\right)  ,\underset{\text{the
number of }J\left(  t\right)  \text{s is }M\left(  t\right)  }{M\left(
t\right)  ,\underbrace{J\left(  t\right)  ,J\left(  t\right)  ,\ldots,J\left(
t\right)  }}\right\}  $ is a level-independent QBD process whose infinitesimal
generator is given by%
\[
\mathbb{Q}=\left(
\begin{array}
[c]{ccccc}%
F_{1} & F_{0} &  &  & \\
F_{2} & A_{1} & A_{0} &  & \\
& A_{2} & A_{1} & A_{0} & \\
&  & \ddots & \ddots & \ddots
\end{array}
\right)  ,
\]
where%
\[
F_{1}=\left(
\begin{array}
[c]{ccccc}%
-\lambda & \lambda\alpha\left(  1\right)  &  &  & \\
T^{0} & T-\lambda I & \lambda D\left(  1\right)  &  & \\
& C\left(  2\right)  & T\left(  2\right)  -\lambda I\left(  2\right)  &
\lambda D\left(  2\right)  & \\
&  & \ddots & \ddots & \ddots\\
&  &  & C\left(  N-1\right)  & T\left(  N-1\right)  -\lambda I\left(
N-1\right)
\end{array}
\right)
\]%
\[
F_{0}=\left(
\begin{array}
[c]{ccccc}
&  &  &  & \\
&  &  &  & \\
&  &  &  & \\
\lambda D\left(  N-1\right)  &  &  &  &
\end{array}
\right)  ,\text{ \ \ }F_{2}=\left(
\begin{array}
[c]{ccccc}
&  &  &  & C\left(  N\right)
\end{array}
\right)  ;
\]%
\[
A_{0}=\underset{\text{Keronecker product of }N\text{ identity matrices}%
}{\lambda\underbrace{I\otimes I\otimes\cdots\otimes I}},\text{\ }%
\]%
\[
A_{1}=\underset{\text{Keronecker sum of }N\text{ matrices }T}{\underbrace
{T\oplus T\oplus\cdots\oplus T}}-\text{ }\underset{\text{Keronecker product of
}N\text{ identity matrices}}{\lambda\underbrace{I\otimes I\otimes\cdots\otimes
I},}%
\]%
\begin{align*}
A_{2} =  &  \left(  T^{0}\alpha\right)  \otimes\underset{\text{Keronecker
product of }N-1\text{ identity matrices}}{\underbrace{I\otimes I\otimes
\cdots\otimes I}}+\text{ }I\otimes\left(  T^{0}\alpha\right)  \otimes
I\otimes\cdots\otimes I+\cdots\\
&  +I\otimes I\otimes\cdots\otimes\left(  T^{0}\alpha\right)  \otimes
I+I\otimes I\otimes I\otimes\cdots\otimes\left(  T^{0}\alpha\right)  .
\end{align*}

Now, we apply the mean drift technique to find a necessary and sufficient
condition under which the QBD process $\mathbb{Q}$ is stable.

\begin{The}
The QBD process $\mathbb{Q}$ is stable if and only if $\rho=\left[
\lambda\left(  \gamma+\mu\right)  \right]  /\left(  N\mu\gamma\right)  <1$.
\end{The}

\textbf{Proof.}\textit{ }We compute%
\begin{align*}
\mathbf{A}=  &  A_{0}+A_{1}+A_{2}\\
=  &  \underset{\text{Keronecker sum of }n\text{ matrices }T}{\underbrace
{T\oplus T\oplus\cdots\oplus T}}+\underset{\text{Keronecker product of
}N-1\text{ identity matrices}}{\left(  T^{0}\alpha\right)  \otimes
\underbrace{I\otimes I\otimes\cdots\otimes I}}\\
&  +\cdots+I\otimes I\otimes I\otimes\cdots\otimes\left(  T^{0}\alpha\right)
\\
=  &  \left(  T+T^{0}\alpha\right)  \otimes I\otimes I\otimes\cdots\otimes
I+\cdots\\
&  +I\otimes I\otimes I\otimes\cdots\otimes\left(  T+T^{0}\alpha\right)  .
\end{align*}
Since the Markov process $T+T^{0}\alpha$ is irreducible and positive
recurrent, it has one unique stationary probability vector $\omega=\left(
\omega_{1},\omega_{2}\right)  $, where%
\[
\omega_{1}=\frac{\mu}{\gamma+\mu},\text{ \ }\omega_{2}=\frac{\gamma}%
{\gamma+\mu}.
\]
It is easy to check that $\Theta=\underset{\text{Keronecker product of
}N\text{ vectors }\omega}{\underbrace{\omega\otimes\omega\otimes\cdots
\otimes\omega}}$ is the stationary probability vector of the the Markov
process $\mathbf{A}$. Also, we get
\[
\Theta A_{2}e=N\mu\frac{\gamma}{\gamma+\mu},\text{ \ }\Theta A_{0}e=\lambda.
\]
By using the mean drift technique, the QBD process $\mathbb{Q}$ is stable if
and only if $\Theta A_{2}e>\Theta A_{0}e$. Note that $\Theta A_{2}e>\Theta
A_{0}e$ is equivalent to%
\[
\rho=\frac{\lambda}{N\mu\frac{\gamma}{\gamma+\mu}}=\frac{\lambda\left(
\gamma+\mu\right)  }{N\mu\gamma}<1.
\]
Therefore, we find a necessary and sufficient condition under which the QBD
process $\mathbb{Q}$ is stable. This completes the proof. \hfill$\blacksquare$

If the QBD process $\mathbb{Q}$ is stable, then it must has a stationary
probabilty vector%
\[
\Psi=\left(  \psi_{0},\psi_{1},\psi_{2},\ldots\right)  ,
\]
where $\psi_{0}$ is related to Level $0$ corresponding to the matrix $B_{1}$,
and $\psi_{k}$ is related to Level $k$ corresponding to the matrix $A_{1}$. To
provide performance evaluation of the second queueing model of service
platforms, we need to express each element of the stationary probabilty vector
$\Psi$ according to the Keronecker sum and the Keronecker product in matrix
computation involved.

It is a little more complicated to express the elements of the vector
$\psi_{0}$. Corresponding to the matrix $B_{1}$ with the Keronecker sums and
the Keronecker products, we have%
\[
\psi_{0}=\left(  \psi_{0,(0)}^{\left(  0\right)  };\psi_{1,(1)}^{\left(
0\right)  },\psi_{1,(2)}^{\left(  0\right)  };\psi_{2,(1,1)}^{\left(
0\right)  },\psi_{2,(1,2)}^{\left(  0\right)  },\psi_{2,(2,1)}^{\left(
0\right)  },\psi_{2,(2,2)}^{\left(  0\right)  };\ldots\right)  .
\]
For $1\leq n\leq N-1$, these elements $\psi_{n,(i_{1},i_{2},i_{3},\ldots
,i_{n})}$ with $i_{k}=1$ or $2$ and for $1\leq k\leq n$ are arranged in a
multi-dimensional lexicographic order. For the multi-dimensional lexicographic
order, we provide two simple examples for $n=2,3$, while for the general case
with $4\leq n\leq N-1$, we can similarly write such a lexicographic order.

Example one: $n=2$. In this case, $\psi_{2,(1,1)}^{\left(  0\right)  }%
,\psi_{2,(1,2)}^{\left(  0\right)  },\psi_{2,(2,1)}^{\left(  0\right)  }%
,\psi_{2,(2,2)}^{\left(  0\right)  }$.

Example two: $n=3$. In this case, $\psi_{3,(1,1,1)}^{\left(  0\right)  }%
,\psi_{3,(1,1,2)}^{\left(  0\right)  },\psi_{3,(1,2,1)}^{\left(  0\right)
},\psi_{3,(1,2,2)}^{\left(  0\right)  },\psi_{3,(2,1,1)}^{\left(  0\right)
},\psi_{3,(2,1,2)}^{\left(  0\right)  }$, $\psi_{3,(2,2,1)}^{\left(  0\right)
},\psi_{3,(2,2,2)}^{\left(  0\right)  }$.

Now, we arrange the elements of the vector $\psi_{l}$ for $l\geq1$. These
elements $\psi_{N,(i_{1},i_{2},i_{3},\ldots,i_{N})}^{\left(  l\right)  }$ with
$i_{k}=1$ or $2$ and for $1\leq k\leq N$ are arranged in the $N$-dimensional
lexicographic order.

Note that $\Psi=\left(  \psi_{0},\psi_{1},\psi_{2},\ldots\right)  $ is the
stationary probability vector of the QBD process $\mathbb{Q}$, then $\Psi$
satisfies the system of linear equations: $\Psi\mathbb{Q}=\mathbf{0\ }$and
$\Psi e=1$.

Let the rate matrix $\mathbf{R}$ be the minimal non-negative solution to the
matrix quadratic equation%
\[
\mathbf{R}^{2}A_{2}+\mathbf{R}A_{1}+A_{0}=0.
\]
By using Chapter 3 of Neuts \cite{Neu:1981}, the stationary probability of the
QBD process $\mathbb{Q}$ is a matrix-geometric solution
\[
\psi_{i}=\psi_{1}\mathbf{R}^{i-1},\text{ }i\geq1,
\]
where $\left(  \psi_{0},\psi_{1}\right)  $ is the unique solution to the
following system of linear equations%
\[
\psi_{0}F_{1}+\psi_{1}F_{2}=0,
\]%
\[
\psi_{0}F_{0}+\psi_{1}\left(  A_{1}+\mathbf{R}A_{2}\right)  =0
\]
and%
\[
\psi_{0}e+\psi_{1}\left(  I-\mathbf{R}\right)  ^{-1}e=1.
\]

In the remainder of this section, by using the stationary probability vector
of the QBD process $\mathbb{Q}$, we provide the performance measures of the
second queueing model of service platforms as follows:

\textbf{(a) The stationary average queue lengths}

Let $\mathcal{Q}^{\left(  1\right)  }$ and $\mathcal{Q}^{\left(  2\right)  }$
denote the stationary queue lengths for the idle owners retained in the
service platform, and for the seekers waiting for their services,
respectively. Then%
\[
E\left[  \mathcal{Q}^{\left(  1\right)  }\right]  =\sum\limits_{n=0}%
^{N-1}\left(  N-n\right)  \sum\limits_{k=1}^{n}\sum_{i_{k}=1,2}\psi
_{n,(i_{1},i_{2},i_{3},\ldots,i_{n})}%
\]
and%
\[
E\left[  \mathcal{Q}^{\left(  2\right)  }\right]  =\sum\limits_{k=1}^{\infty
}k\Psi_{k}e=\Psi_{1}\left(  I-R\right)  ^{-2}\mathbf{e}.
\]

\textbf{(b) The expected profits of the service platform and each owner}

Let $f_{1}$ and $f_{2}$ denote the expected profits per unit time of a service
platform and each owner, respectively. Then%
\[
f_{1}=\left(  1-d\right)  \cdot P\left(  N-E\left[  \mathcal{Q}^{\left(
1\right)  }\right]  \right)  \mu,
\]
and%
\[
f_{2}=d\cdot P\left(  1-\frac{1}{N}E\left[  \mathcal{Q}^{\left(  1\right)
}\right]  \right)  \mu.
\]

\section{Numerical Examples}

In this section, we uses numerical examples to discuss how the performance
measures depend on some key system parameters. Note that the numerical
analysis is useful and necessary in the design and operations management of
the service platforms.

Note that the number of owners $N$ is a key factor that determines the sizes
of blocks (for example, the matrices $A$, $B$ and $C$) in the infinitesimal
generator $Q$. Also, the sizes of blocks directly affect algorithm design and
computational complexity of the level-independent QBD process $Q$. Obviously,
the larger the number $N$, the more difficult the numerical computation is.

In the first two parts (a) and (b), the selected parameters $\lambda$, $\mu$,
$\gamma$ and $N$ must satisfy the system stable condition: $\rho
=\lambda\left(  1+\mu/\gamma\right)  /N\mu<1$.

\textbf{(a) The stationary average queue lengths}

To obtain the stationary probability vector, we first need to compute the rate
matrix $R$, which is the minimal nonnegative solution to the quadratic
nonlinear matrix equation: $R^{2}A+RB+C=0$. A modified successive iteration
method is found in, e.g., see Neuts \cite{Neu:1981} and Latouche and Ramaswami
\cite{Lat1999}. Now, we describe the modified successive iteration method as
follows. The sequence $\left\{  R\left(  n\right)  ;n\geq0\right\}  $ is
designed as%
\[
R\left(  0\right)  =\mathbf{0,}%
\]%
\[
R\left(  n+1\right)  =-\left[  R\left(  n\right)  ^{2}A+C\right]
B^{-1},\text{ \ \ }n=0,1,2,\ldots.
\]
Neuts \cite{Neu:1981} indicated that $R\left(  n\right)  \uparrow R$ as
$n\rightarrow\infty$. For any sufficiently small positive number $\varepsilon$
within the desired degree of accuracy, set at $10^{-12}$, if there exists a
positive integer $n$ such that
\[
\left\|  R\left(  n+1\right)  -R\left(  n\right)  \right\|  =\max\left|
R\left(  n+1\right)  _{i,j}-R\left(  n\right)  _{i,j}\right|  <\varepsilon,
\]
then we take $R=R\left(  n\right)  $.

Once the rate matrix $R$ is computed numerically, the two vector $\pi_{0}$ and
$\pi_{1}$ can be obtained by solving Equations (\ref{equ2}), (\ref{equ3}),
(\ref{equ5}), and (\ref{equ6}). Further, by substituting $\pi_{1}$ into
Equation (\ref{equ6}), we can obtain $\pi_{n}$ for $n\geq2$.

\textit{(1) The effect of the arrival rate }$\lambda$\textit{:} We take the
parameters: $N=60$, $\mu=1$, $\gamma=100$, and $\lambda\in\left[
10,46\right]  $.

The left of Figure 4 shows that $E\left[  \mathcal{Q}^{\left(  1\right)
}\right]  $ decreases and $E\left[  \mathcal{Q}^{\left(  2\right)  }\right]  $
increases as the arrival rate $\lambda$ increases. Such two numerical results
are intuitive from our practical observation. As the arrival rate $\lambda$
increases, the seekers arrive at the service platform at a faster pace. This
leads to the result that fewer idle owners are retained in the service
platform, while more seekers have to wait for their services.

\textit{(2) The effect of the matching rate }$\gamma$\textit{:}\textbf{ }We
take the parameters: $N=60$, $\lambda=10$, $\mu=1$, and $\gamma\in\left[
100,300\right]  $.

The right of Figure 4 indicates that both $E\left[  \mathcal{Q}^{\left(
1\right)  }\right]  $ and $E\left[  \mathcal{Q}^{\left(  2\right)  }\right]  $
decrease as the matching rate $\gamma$ increases. When the matching rate
$\gamma$ increases, the owners enter service state faster, and, hence, the
seekers receive their services and leave the system more quickly.

\begin{figure}[th]
\setlength{\abovecaptionskip}{0.cm}  \setlength{\belowcaptionskip}{-0.cm}
\centering                         \includegraphics[width=7cm]{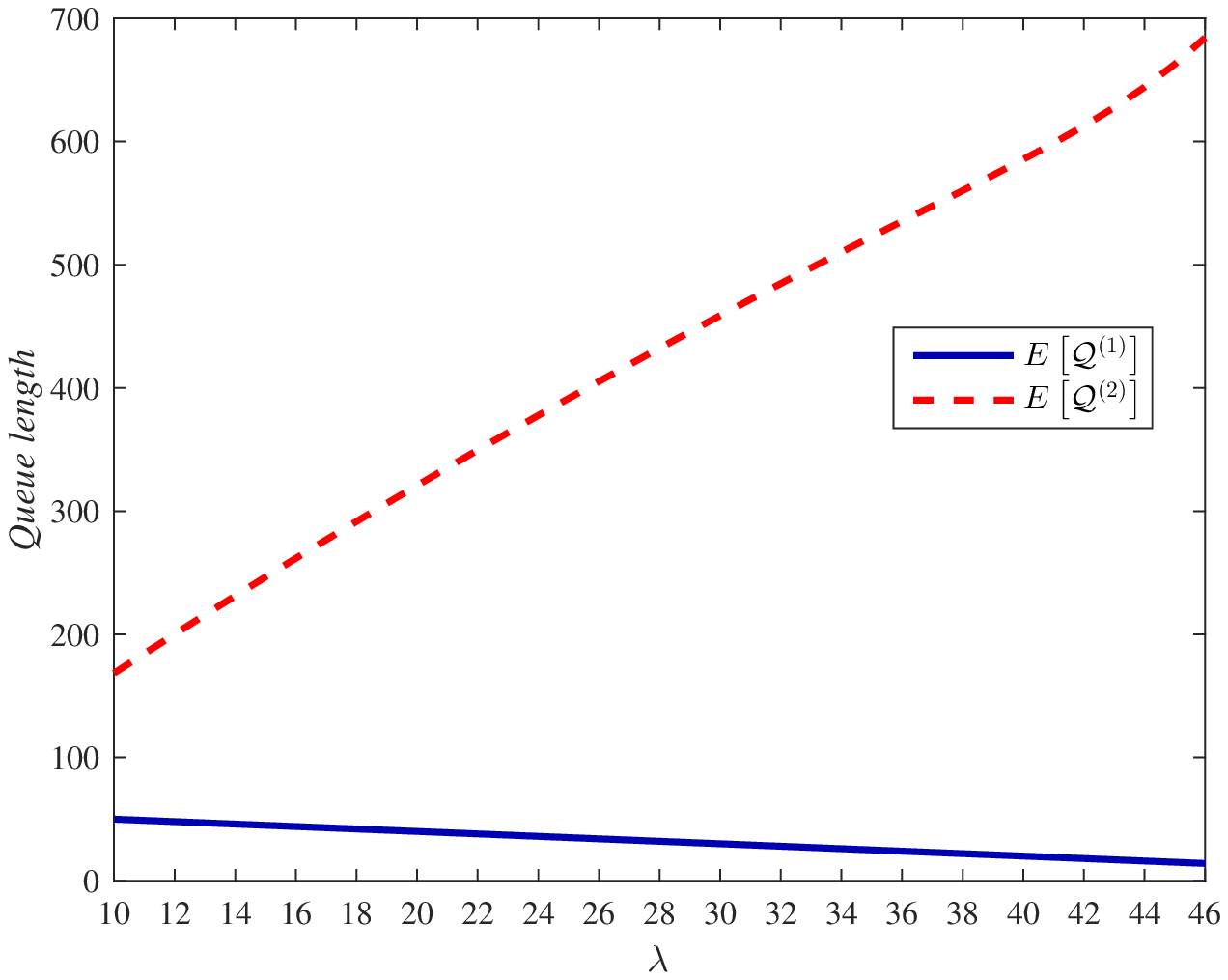}
\includegraphics[width=7cm]{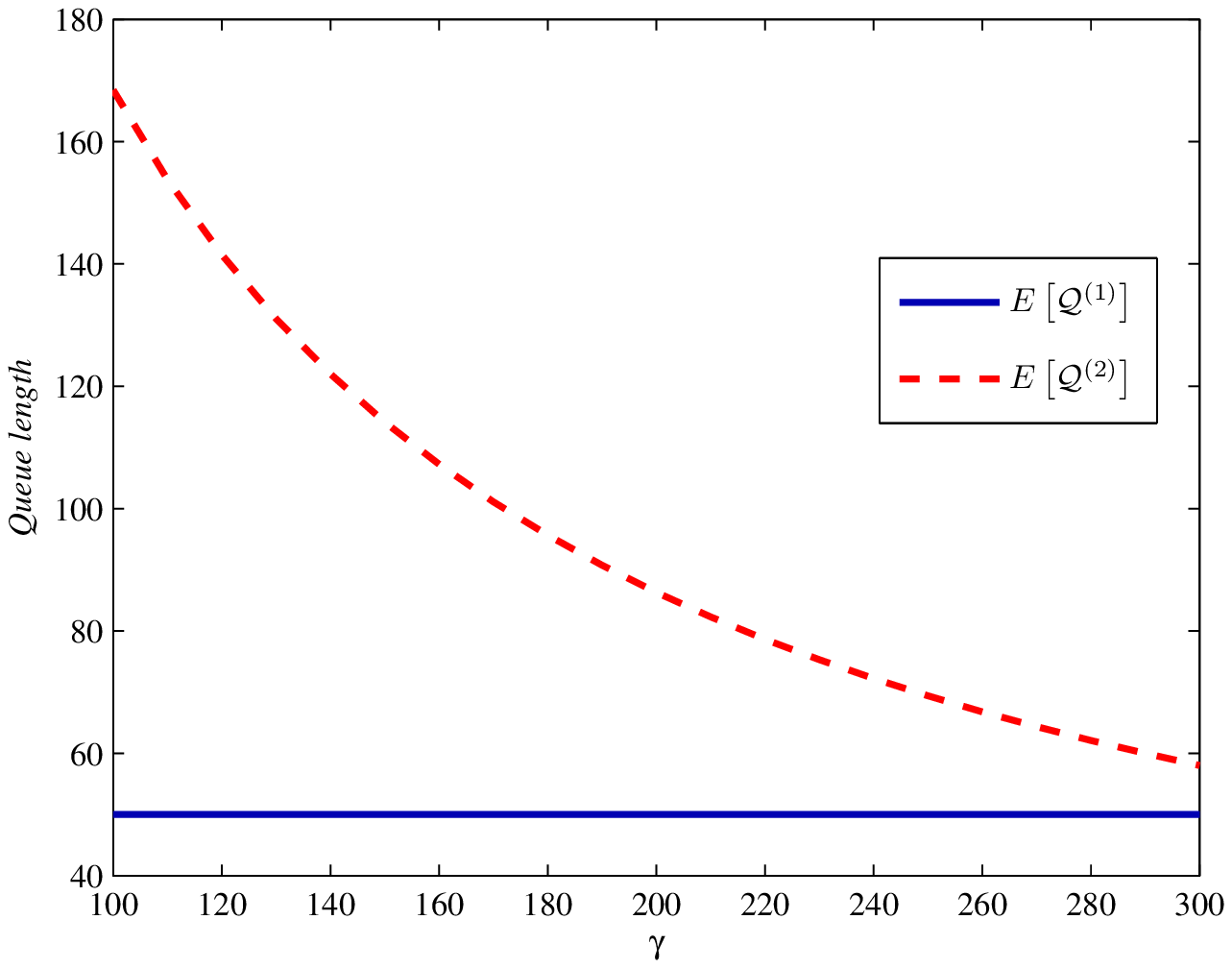}  \newline \caption{$E\left[
\mathcal{Q}^{\left(  1\right)  }\right]  $ and $E\left[  \mathcal{Q}^{\left(
2\right)  }\right]  $ vs. $\lambda$ and $\gamma$}%
\label{figure:figure-4}%
\end{figure}

\textit{(3) The effect of the owner number }$N$\textit{:}\textbf{ }We take the
parameters: $\lambda=10$, $\mu=0.26$, $\gamma=100$, and $N=43,44,45,\ldots,53$.

Figure 5 shows that $E\left[  \mathcal{Q}^{\left(  1\right)  }\right]  $
increases and $E\left[  \mathcal{Q}^{\left(  2\right)  }\right]  $ decreases
as the owner number $N$ increases. When the owner number $N$ increases, more
idle owners are retained in the service platform, while more seekers can
receive their services and leave the system immediately. Thus, fewer seekers
are kept waiting.

\begin{figure}[tbh]
\centering            \includegraphics[height=5.5cm
,width=10cm]{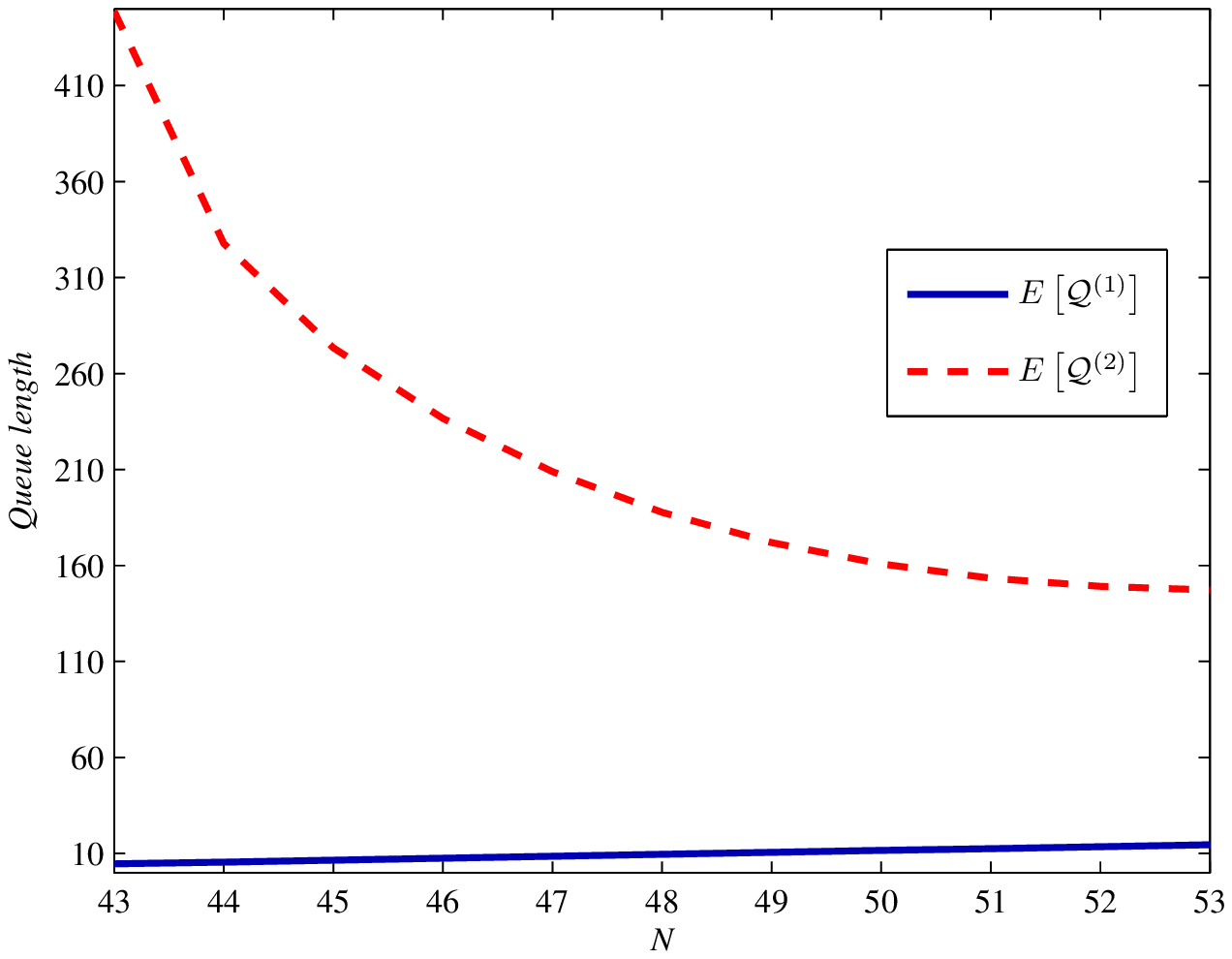}
\caption{Effect of $N$ on $E\left[  \mathcal{Q}^{\left(  1\right)  }\right]  $
and $E\left[  \mathcal{Q}^{\left(  2\right)  }\right]  $}%
\end{figure}

\textbf{(b) The expected profits for the service platform and each owner}

\textit{(1) The effect of the arrival rate }$\lambda$\textit{: }We take the
parameters: $N=60$, $\mu=1$, $\gamma=100$, $P=50$, $d=0.8$ and $\lambda
\in\left[  10,46\right]  $.

The left of Figure 6 that $f_{1}$ and $f_{2}$ increase as the arrival rate
$\lambda$ increases. Such numerical results are consistent with our intuitive
understanding. As the arrival rate $\lambda$ increases, more seekers enter the
system per unit time, which directly improves the expected profits for both
the service platform and each owner.

\textit{(2) The effect of the service price }$P$\textit{:} We take the
parameters: $N=60$, $\lambda=10$, $\mu=1$, $\gamma=100$, $d=0.8$, and
$P\in\left[  30,50\right]  $.

From The right of Figure 4, we show that $f_{1}$ and $f_{2}$ increase as the
service price $P$ increases. Clearly, the service platform and each owner earn
more with a higher service price $P$.

\begin{figure}[th]
\setlength{\abovecaptionskip}{0.cm}  \setlength{\belowcaptionskip}{-0.cm}
\centering                         \includegraphics[width=7cm]{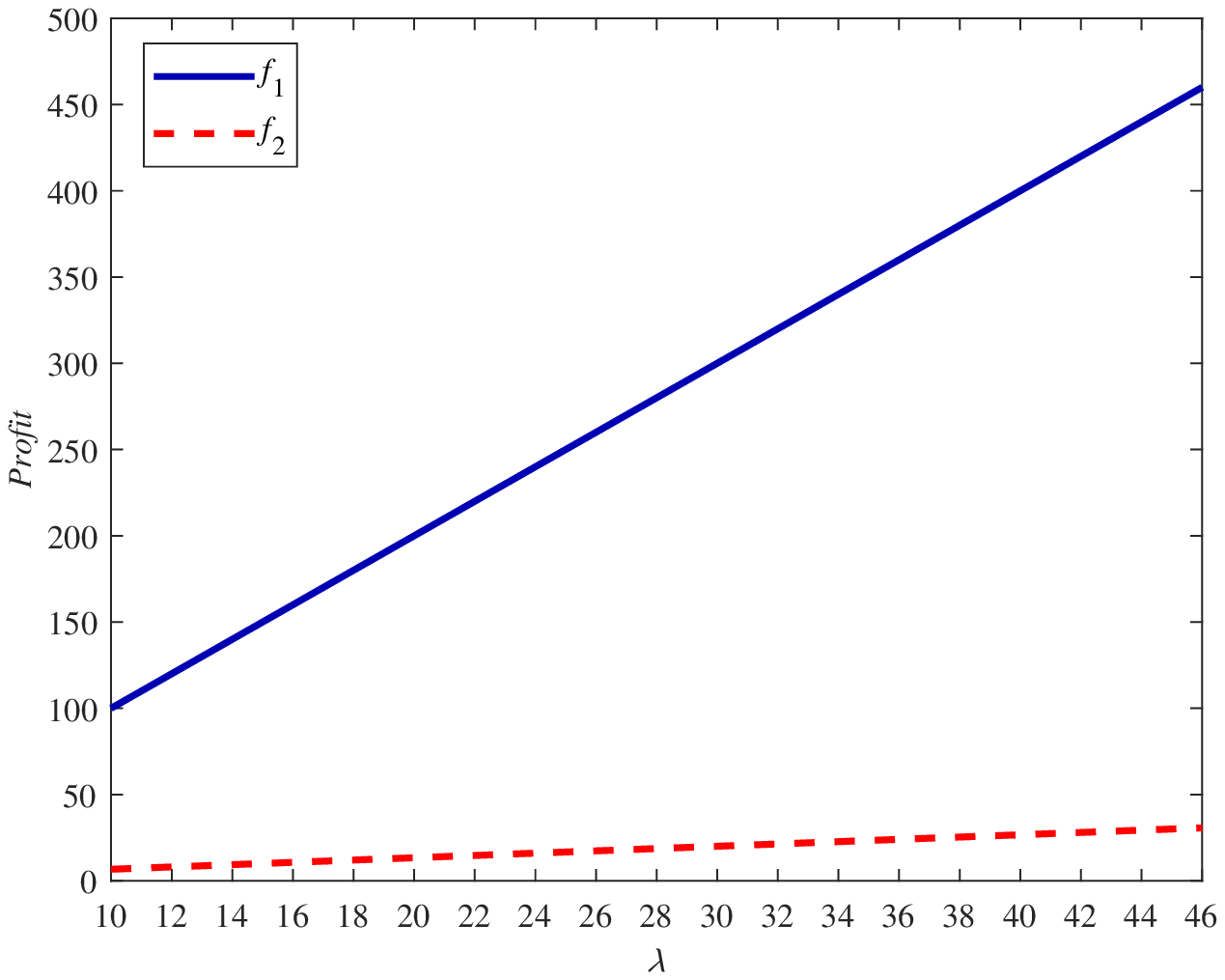}
\includegraphics[width=7cm]{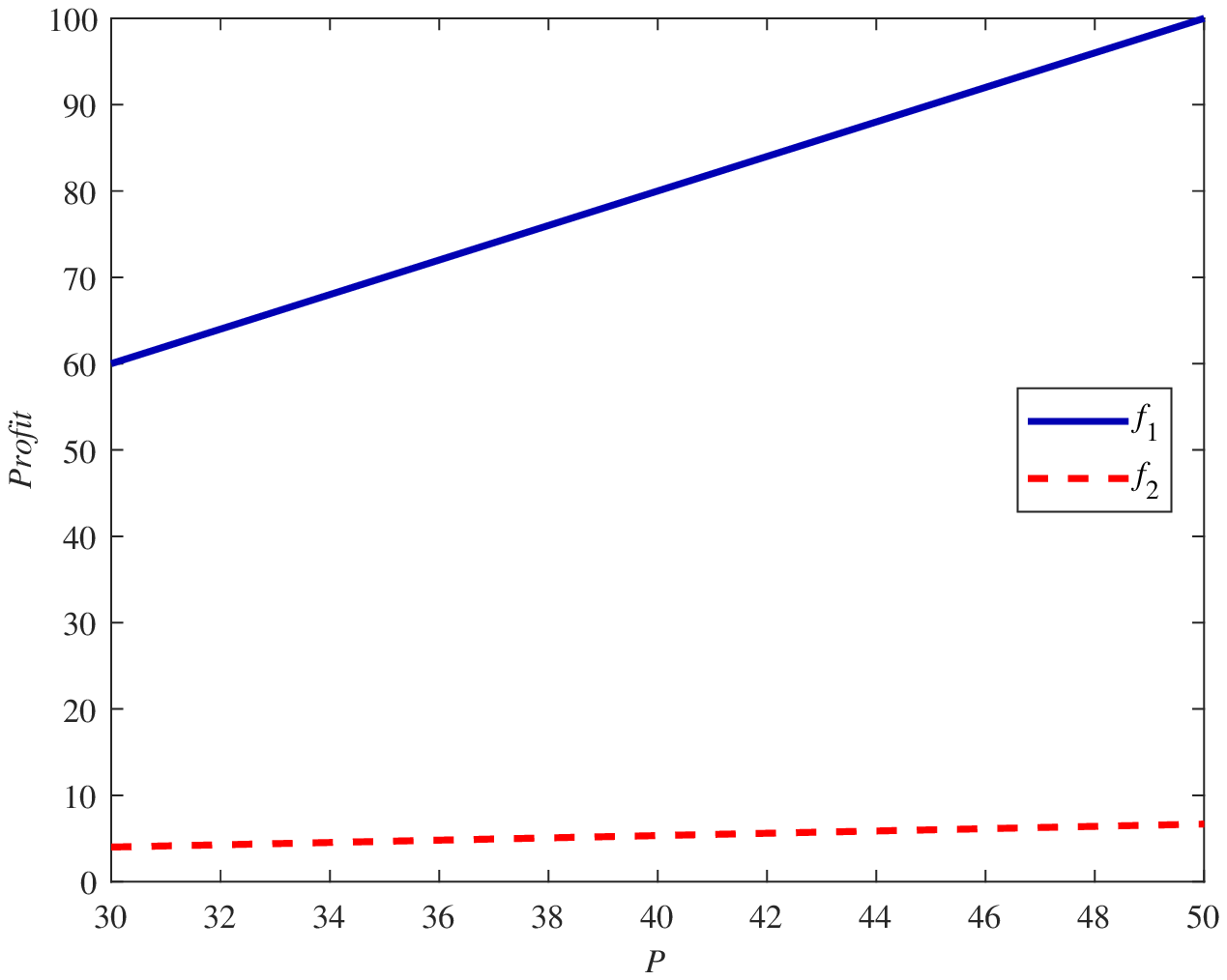}  \newline \caption{Effect of
$\lambda$ and $P$ on $f_{1}$ and $f_{2}$}%
\label{figure:figure-6}%
\end{figure}

\textbf{(c) The expected sojourn time}

\textit{(1) The effect of the arrival rate }$\lambda$\textit{:} We take the
parameters: $N=60$, $\mu=1$, $\gamma=100$, $P=50$, $d=0.8$, and $\lambda
\in\left[  10,46\right]  $.

The left of Figure 7 shows that the expected sojourn time $E\left[  W\right]
$ increases as the arrival rate $\lambda$ increases. The numerical result is
intuitive. With a higher arrival rate, more seekers wait for their services.
This leads to a larger expected sojourn time.

\textit{(2) The effect of the number of owners }$N$\textit{:} We take the
parameters: $\lambda=10$, $\mu=0.26$, $\gamma=100$, $P=50$, $d=0.8$, and
$N=43,44,45,\ldots,53$.

The rightt of Figure 7 shows that the expected sojourn time $E\left[
W\right]  $ decreases as $N$ increases. When the number of owners $N$
increases, the seekers can be more quickly matched with the owners so that the
expected sojourn time decreases.

\begin{figure}[th]
\setlength{\abovecaptionskip}{0.cm}  \setlength{\belowcaptionskip}{-0.cm}
\centering                         \includegraphics[width=7cm]{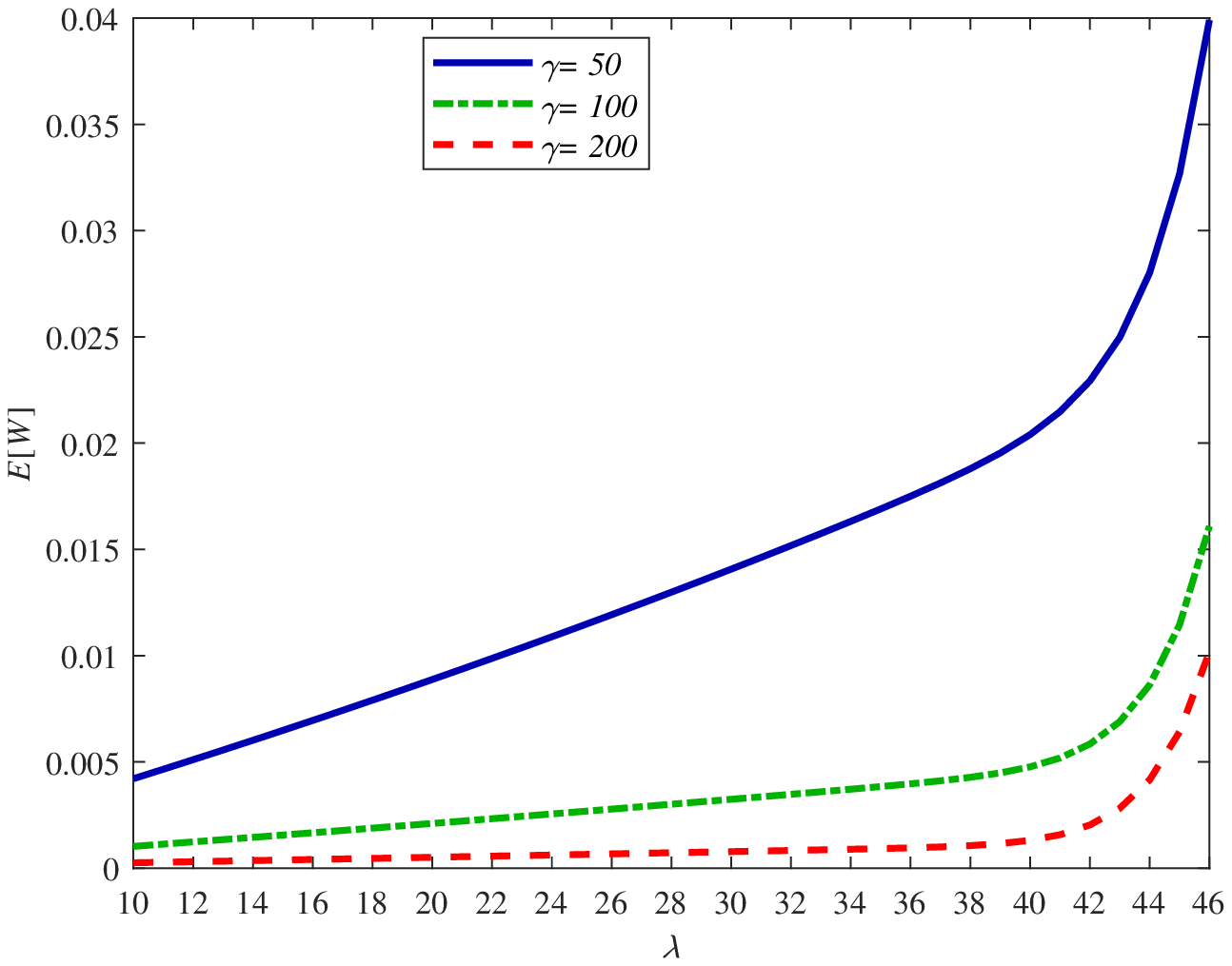}
\includegraphics[width=7cm]{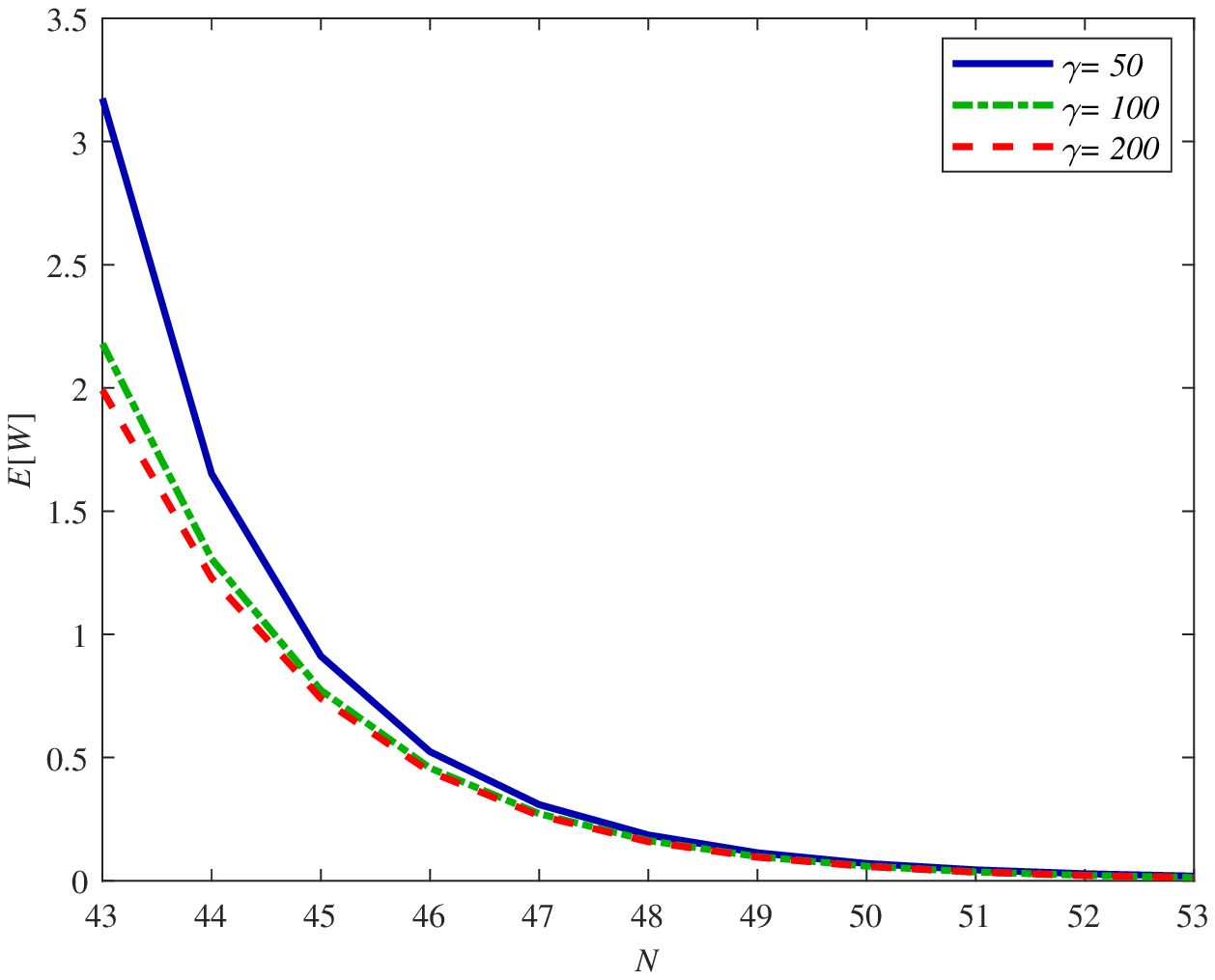}  \newline \caption{Effect of
$\lambda$ and $N$ on $E[W]$}%
\label{figure:figure-7}%
\end{figure}

\section{Concluding remarks}

In this paper, we firstly describe two basic queueing models of service
platforms in digital sharing economy according to two different policies of
platform matching information: One is at the matching completion time (it is
also the service beginning time), while another is at the matching beginning
time (no service yet). Then we express the two basic queueing models of
service platforms as the level-independent QBD processes, and apply the
matrix-geometric method to obtain a necessary and sufficient condition under
which the system is stable, the stationary probability vector, the expected
sojourn time by using the RG-factorizations, and the expected profits per unit
time of a service platform and each owner. This enables performance evaluation
of the service platforms. Finally, we use some numerical examples to indicate
how the performance measures are influenced by some key system parameters.

We believe that the methodology and results given in this paper are applicable
to more extensive queueing analysis of service platforms, and will open a
series of promising research, such as in the following directions:

- Extending our basic queueing models to the Markovian arrival process of
seekers, and to that one owner can match more than one seekers as a pair.

- Considering several types of owners and/or seekers, and introducing
different matching and service priorities, and full-time and part-time owners.

- Based on our two basic queueing models, developing Markov decision processes
of service platforms.

- Setting up queueing-game to analyze the owners' and/or seekers' strategic
behaviors in the study of service platforms and digital sharing economy.

\section*{Acknowledgements}

Quan-Lin Li was supported by the National Natural Science Foundation of China
under grants No. 71671158 and 71932002.

\end{document}